\documentclass[preprint,12pt,authoryear]{article}
\usepackage{lineno,hyperref}
\usepackage{ragged2e}
\usepackage{array}
\usepackage{multirow}
\usepackage[margin=1in]{geometry}
\usepackage{graphicx,subfigure}
\usepackage{multirow}
\usepackage{epstopdf}
\usepackage{setspace}
\usepackage{scalerel}
\usepackage{bbold}
\usepackage{float}
\usepackage{amsmath, amssymb, ulem}
\usepackage{rotating}
\usepackage{morefloats}
\usepackage{stackengine}
\usepackage{comment}
\usepackage{caption}
\usepackage{natbib}
\setcitestyle{round}
\usepackage{fancybox,color}
\usepackage{booktabs}
\usepackage[utf8]{inputenc}
\usepackage[english]{babel}
\usepackage{subcaption}
\usepackage{tabularx}
\newlength{\Oldarrayrulewidth}
\newcommand{\Cline}[2]{%
  \noalign{\global\setlength{\Oldarrayrulewidth}{\arrayrulewidth}}%
  \noalign{\global\setlength{\arrayrulewidth}{#1}}\cline{#2}%
  \noalign{\global\setlength{\arrayrulewidth}{\Oldarrayrulewidth}}}

\makeatletter 
\renewcommand\@biblabel[1]{} 
\makeatother

\usepackage[margin=1in]{geometry}
\usepackage{amsmath,amsthm,amssymb,amsfonts,graphicx,dsfont}

\usepackage{booktabs,parskip}
\usepackage{epsfig}

\usepackage{url}
\usepackage{bm}
\usepackage{rotating}
\usepackage{listings}
\usepackage{verbatim}
\usepackage{xcolor}
\usepackage[title]{appendix}
\usepackage{multirow}

\usepackage[ruled,vlined,linesnumbered]{algorithm2e}
\SetKwFunction{KwFn}{Fn}
\SetKwInOut{Input}{input}
\SetKwInOut{Output}{output}

\usepackage{longtable}
\usepackage{adjustbox}
\usepackage{subfigure}
\hypersetup{colorlinks,%
citecolor=black,%
filecolor=black,%
linkcolor=black,%
urlcolor=black
}

\usepackage{setspace} 

\newcommand{\eval}[2][\right]{\relax
	\ifx#1\right\relax \left.\fi#2#1\rvert}

\newtheoremstyle{break}
  {\topsep}{\topsep}%
  {\itshape}{}%
  {\bfseries}{}%
  {\newline}{}%
\theoremstyle{break}

\newtheorem{proposition}{Proposition}
\theoremstyle{definition}

\newtheorem{exampleemph}[proposition]{Example}   

\usepackage{changepage}   
\makeatletter
\newcommand*{\rom}[1]{\expandafter\@slowromancap\romannumeral #1@}
\makeatother

\title{European Option Pricing in Regime Switching Framework via Physics-Informed Residual Learning}
\author{Naman Krishna Pande \and Puneet Pasricha \and Arun Kumar \and Arvind Kumar Gupta}

\date{Department of Mathematics, Indian Institute of Technology Ropar, Punjab-140001, India}

\begin{document}
\maketitle
\begin{abstract}
In this article, we employ physics-informed residual learning (PIRL) and propose a pricing method for European options under a regime-switching framework, where closed-form solutions are not available. We demonstrate that the proposed approach serves an efficient alternative to competing pricing techniques for regime-switching models in the literature. Specifically, we demonstrate that PIRLs eliminate the need for retraining and become nearly instantaneous once trained, thus, offering an efficient and flexible tool for pricing options across a broad range of specifications and parameters.
\end{abstract}

\textbf{Keywords:} European Options; Regime-Switching; Physics-Informed Residual Learning; Option Pricing; Numerical Solution; Partial Differential Equations (PDE).

\section{Introduction}

Option pricing is one of the central problems in mathematical finance, both from theoretical and practical perspectives, as investors use options not only as a speculation instrument but also as a hedging tool against their investment risk. Options are derivative products that give its holder the right but not the obligation to buy or sell the underlying asset subject to certain conditions. Ever since the ground-breaking work of Black, Scholes and Merton, i.e. the Black-Scholes Merton (BSM) option pricing model in 1973, numerous studies have adopted various stochastic models to overcome the shortcomings of BSM to capture the dynamics of the underlying assets more accurately. For instance, stochastic volatility models are typically adopted to address the volatility smile, mean-reversion, fat tails, and volatility clustering, while the jump-diffusion models are introduced to describe the jumps and their clustering behavior. The regime-switching models have also become popular as the governing continuous time Markov chain allows for influence of major economic factors on asset price dynamics in a parsimonious manner. For instance, it allows the volatility parameter to switch based on the state of underlying Markov chain. 

Various numerical techniques have been proposed to obtain the approximate solutions since the closed-form formulae for option prices under these advanced dynamics are generally unavailable or not easily derived. In the finance industry, the most commonly used approaches to price derivatives are Monte-Carlo (MC) simulations, binomial trees, solving partial differential equations or inversion of the characteristic function (Fast Fourier Transform (FFT)). However, in the current scenario, one cannot rely on the above-mentioned time-consuming techniques primarily due to the introduction of more complex derivative products and, hence, complex models to price these derivatives. More specifically, in derivative markets, countless computations are performed daily, such as calibration of models to market quotes, pricing derivatives, calculating hedge positions, determining risk management indicators, etc. The information from these calculations is often useful only for a limited time. As time progresses and the market fluctuates, these values quickly become outdated and necessitate updates. Indeed, real-time updates and information are vital for the successful operation of a derivatives business. As a result, the need to develop eﬃcient pricing methods to obtain option values becomes ever more prominent for practitioners and academics alike.

Over the recent past, machine learning has experienced a remarkable evolution, which can be attributed to rapid growth in computation and storage capacity, availability of extensive data sets, etc. On one hand, the development of powerful learning techniques and their applications in computer science and engineering are omnipresent. On the other hand, quantitative finance has only recently started seeing applications of these innovative and powerful machine learning (ML) methods (see \cite{filipovic2022empirical, de2018machine,anderson2023accelerated,mikkila2023empirical}). One of the recent developments in ML, has led to the invention of a rather new class of models, known as physics-informed deep learning (PIDL), that fuses together the representation power of the ML and the physical prowess of theoretically proven mathematical models (\cite{raissi}) into a single learning framework. These models augment the physical laws into the architecture of the ML framework, in the form of a (system of) partial differential equation(s), thereby generating a robust predictive model that aligns with the governing properties of the system under consideration. The physical laws act as a guidance system for the ML framework by restricting the space of admissible solutions to the target domain. This allows for more generalisable solutions and reduces the data requirements of training by a significant amount. Since its conception, PIDL has been applied in solving various forward and inverse problems arising in the domain of partial differential equations like, Schrodinger's equation, Burger's equation, Allen Cahn equations (\cite{raissi}). Some variants of PIDLs like, Galerkin method based hp-variational physics-informed neural network (\cite{kharazmi2021hp}), Bayesian PIDL (\cite{yang2021b}), physics-informed adversarial neural networks (\cite{shekarpaz2022piat}), have been proposed for forward and inverse problems in partial differential equations. \cite{CHEN2024128480} proposed a physics-informed data-driven algorithm for ensemble forecast of complex turbulent system, \cite{ wang2023deep} utilised the convolution neural network based physics-informed Residual Network for rolling element bearing fault diagnostics. PIDLs have also been applied to solve the traffic flow problem by utilising the macroscopic traffic flow models, i.e the LWR model (\cite{10105558, pidlTSE}) and ARZ model (\cite{Shi_Mo_Di_2021, arzfdl}) and hybrid models combining the physical aspects of cell transmission model with the deep neural network framework (\cite{lwrctm}). Recently, PIDL framework for pricing the derivatives has garnered interest. \cite{wang2023deep} developed a PIDL framework for solving the BSM equation and showed the efficacy of the PIDL based pricing methods over traditional numerical schemes. \cite{bae2024option} utilised the PIDL framework for pricing the option and estimating local volatility surface under the BSM model with constant elasticity of variance. \cite{hainaut2024option} utilised the PIDL framework for pricing the put options under the Heston stochastic volatility model. \cite{hainaut2023valuation} devised a PIDL based framework for the valuation of contracts related to guaranteed minimum accumulation benefits which protects the policyholders against downside market risks. 

In this article, we consider problem of valuation of European options in the presence of stochastic volatility in a regime-switching scenario. More specificially, we demonstrate that the physics-informed residual learning (PIRL) model, is able to efficiently generate the option prices for a wide array of parameters and specifications. In essence, the efficiency of the DL models in solving various scientific problems is enhanced by incorporating the physical laws of the mathematical option pricing framework in its learning process. In our regime-switching framework, the driving principle is the Feynman-Kac (FK) theorem, resulting in a coupled partial differential equation.

Our paper makes several contributions. First, we apply a variant of a standard deep neural network model, i.e. the PIRL network, for pricing the European put options in regime switching economy governed by a system of coupled partial differential equations. This has not yet been explored in the literature upto the best of our knowledge. Second, we conduct experiments on two of the most important option pricing models namely, the BSM model and Heston stochastic volatility model under the framework of regime switching. The models that we consider generalises some of the existing recent works in the literature on application of PIDL in option pricing. Finally, we conduct extensive experiments to test the efficiency of PIDL in learning the prices of the European put options under varying market conditions. Specifically, we train  physics informed residual learning framework for a wide array model parameters, which once trained become nearly instantaneous. Thus the training of the learning framework allows swift computation of option prices for any given configuration of parameters compared to classical pricing methods which require re-computations whenever parameters are altered.


The outlines of this article are as follows: The next section introduces mathematical modelling frameworks for pricing the options. In section 3, a brief overview of the proposed PIRL network for pricing the put options is presented. Section 4 establishes the calibration and training procedure for PIRL network. Section 5 gives a detailed numerical analysis of the results and the last section concludes the work.

\section{Modeling Framework}

Consider a frictionless financial equity market where we model the economy's uncertainty by a filtered probability space $(\Omega,\mathcal{F},\mathbb{P},\mathcal{F}_{t\in[0,T]})$.
Further, we assume that the market is composed of two primary assets, (a) a cash account that accrues interest at a constant risk-free rate $r$, and (b) an underlying asset with price at time $t$ denoted by $S_t$. To capture the impact of different states of the economy, we subject the parameters in the dynamics of $S_t$ to random shifts between different regimes of the economy, modelled by a stochastic process $X_t$. We, further, assume that $X_t$ is a continuous time Markov chain (CTMC) with two states,\footnote{We assume two states for illustration purposes, however, it can be extended to arbitrary but finite states.}
\[
X_t = \begin{cases}
1, & \text{when the economy is regarded as being in State 1, say, the bullish regime;} \\
2, & \text{when the economy is regarded as being in State 2, say, the bearish regime.}
\end{cases}
\]
Following Elliott et al. (2008), we can consider the state space of \( X_t \) as the set of unit vectors \( \{ e_1, e_2 \} \), with \( e_1 = (1, 0)^T \) and \( e_2 = (0, 1)^T \) where $^T$ denotes the transpose of the vector. Since $X_t$ is a CTMC, the transition between the two states occurs as a Poisson process, i.e.,
\begin{equation}
    \mathbb{P}(t_{jk}^* > t) = e^{-\lambda_{jk} t}, \quad j, k = 1, 2, \; j \neq k,
\end{equation}
where $\lambda_{jk}$ is the rate of transition from state $j$ to state $k$ and $t_{jk}^*$ is the time taken by the process in state $j$ to enter the state $k$. In this paper, we consider pricing a European financial derivative with maturity time $T$ and payoff function $H(S_T)$. The fair value of the derivative is given by the expected discounted value of the payoff under the risk-neutral measure
\begin{equation*}
P_t=\mathbb{E}^Q(e^{-r(T-t)}H(S_T)).    
\end{equation*}
To employ PIRL network for pricing European options under a regime-switching scenario, we consider two different modeling frameworks. Further, we consider a put option with a strike price $E$, whose payoff has the following form
\begin{equation*}
H(S_T) = (E -S_T )^+ .
\end{equation*}

\subsection{Black-Scholes model with regime-switching}
\label{bsrs}

Under the physical measure $\mathbb{P}$, we assume that the dynamics of $S_t$ is governed by the stochastic differential equation
\begin{equation}
    dS_t = \mu_{X_t} S_t dt + \sigma_{X_t} S_t dW_t,
\end{equation}
where we assume that the drift rate, $\mu_{X_t}$, and the volatility rate, $\sigma_{X_t}$, are subject to regime-shifts modelled by $X_t$. More specifically $\mu_{X_t}=<\mu,X_t>$ where $\mu=(\mu_1,\mu_2)$ is a vector with $\mu_i$ denoting the value of $r$ when $X_t$ is in state $i$. Similarly, we have $\sigma_{X_t}=<\sigma,X_t>$ where $\sigma=(\sigma_1,\sigma_2)$ such that $\sigma_1\neq \sigma_2$, i.e., $\sigma_{X_s} \neq \sigma_{X_t}$ if $X_s \neq X_t$. Further, the process $\{W_t\}$ is the standard Wiener process and we assume that the processes $X$ and $W$ are independent. 

We need to find an equivalent martingale measure to price a European option. We observe that the price of risk corresponding to a state transition cannot be determined uniquely. Following \cite{zhu2012new}, we make the assumption that the risk related to a state transition is diversifiable and therefore unpriced, an assumption that does not compromise generality (refer \cite{naik1993option}). Following \cite{buffington2002regime}, the system of coupled Black-Scholes equations governing the price of European option is given by
\begin{equation}
    \begin{split}
        &\partial_tV_i  = D_i[V_1, V_2, r, \sigma_1, \sigma_2],\; i=1,\;2, t\in[0,T];\\
        &V_i(0,t) = Ee^{-r(T-t)};\\
        &V_i(S,T) = \max(E-S, 0);\\
        &\lim_{S \to \infty} V_i(S,t)  = 0,\\
    \end{split}
\end{equation}
where $D_i = -\frac{1}{2}\sigma_i^2S^2 \frac{\partial^2 V_i}{\partial S^2} - rS \frac{\partial V_i}{\partial S} + rV_i +\lambda_{ij}(V_i - V_j),\;i\neq j,\;i,\;j=1,2$,  $V_i(S, t) (i = 1, 2)$ is the option value when the economy is in state $i$, $S$ is the value of the underlying asset, $t$ is the current time, $E$ is the strike price and $T$ is the expiration time of the option. Note that when the two transition rates \( \lambda_{12} \) and \( \lambda_{21} \) are zero, we obtain the standard Black-Scholes model. The same holds true when we choose $\sigma_1=\sigma_2$, i.e., the volatility is constant irrespective of the regime in the economy.

\subsection{Heston model with regime-switching}

Under the risk-neutral measure, we assume that the dynamics of the underlying asset is governed by the stochastic differential equation\footnote{\cite{he2016analytical} considered this model and derived an analytical approximation pricing formula for the European put options.}, 
\begin{equation*}
dS_t = rS_tdt + \sqrt{v_t}S_tdW_t^1, \end{equation*}
where $v_t$ is the stochastic variance process governed by the following regime switching stochastic differential equation
\begin{equation}
dv_t = k(\gamma - v_t) \, dt + \sigma_{X_t} \sqrt{v_t} \, dW_t^2, \tag{2.1}
\end{equation}
where $k, \gamma>0$ are the mean-reverting rate and level, respectively. Further, $\sigma_{X_t}$ is the volatility of volatility subject to regime switches modeled by $X_t$ as in the case of Black-Scholes model discussed previously. Furthermore, the processes $W_t^1$ and $W_t^2$ are two standard Brownian motions with correlation \(\rho \) and are assumed to be independent of the Markov chain $\{X_t\}$.

Applying the Itô lemma to \(V\) gives the following partial differential equations

\begin{equation*}
dV = \frac{\partial V}{\partial t} dt + \frac{\partial V}{\partial v} dv + \frac{\partial V}{\partial S} dS + \frac{1}{2} \frac{\partial^2 V}{\partial S^2} (dS)^2 + \frac{1}{2} \frac{\partial^2 V}{\partial v^2} (dv)^2 + \frac{\partial^2 V}{\partial S \partial v} dS dv + \langle (V_1,V_2), dX_t \rangle,
\end{equation*}
where $V_i$ is the price $V$ in state $i$ of the economy. Since the discounted option price should be a martingale, we have the following coupled PDE system governing $V$,

\begin{equation}
    \begin{split}
        & \partial_tV_i  = D_i[V_1, V_2, r, \sigma_1, \sigma_2, \kappa, \gamma, \rho],\; i=1,\;2, t\in[0,T];\\
        & V_i(0, t) = Ee^{-r(T-t)};\\
        & V_i(S, T) = \rm{max}(E-S, 0);\\
        & \lim_{S \to \infty} V_i(S,t)  = 0;\\
        & \lim_{v\to\infty}V_i(S,v,t) = 0;\\
        & \lim_{v\to 0}V_i(S,v,t)=\max\{0, Ee^{-r(T-t)-S}\},
    \end{split}
    \label{hestonpde}
\end{equation}
where $D_i = -\frac{1}{2} v S^2 \frac{\partial^2 V_i}{\partial S^2} - r S \frac{\partial V_i}{\partial S} + r V_i - \frac{1}{2} \sigma_1^2 v \frac{\partial^2 V_i}{\partial v^2} - \sigma_1 v \rho S \frac{\partial^2 V_i}{\partial S \partial v} - k (\gamma - v) \frac{\partial V_i}{\partial v} + \lambda_{ij} (V_i - V_j),\;i\neq j,\;i,\;j = 1, 2$, $S$ is the value of the underlying asset, $E$ is the strike price and $T$ is the time to maturity of the option. Note that when $\lambda_{12}=\lambda_{21}=0$, the model reduces  
to the standard Heston model, under which the pricing of European option was recently studied by \cite{hainaut2024option} using physics-inspired neural networks (PINN). 

\section{Physics informed residual learning}
In this section a brief overview of the PIRL network is presented that is utilised to approximate the price of the European put options in two different economic states. The methodology is generic and can easily be generalised for valuing the European option prices in $n$ different economic states.

\begin{figure}[h]
    \centering
    \includegraphics[width=15cm, height = 8cm]{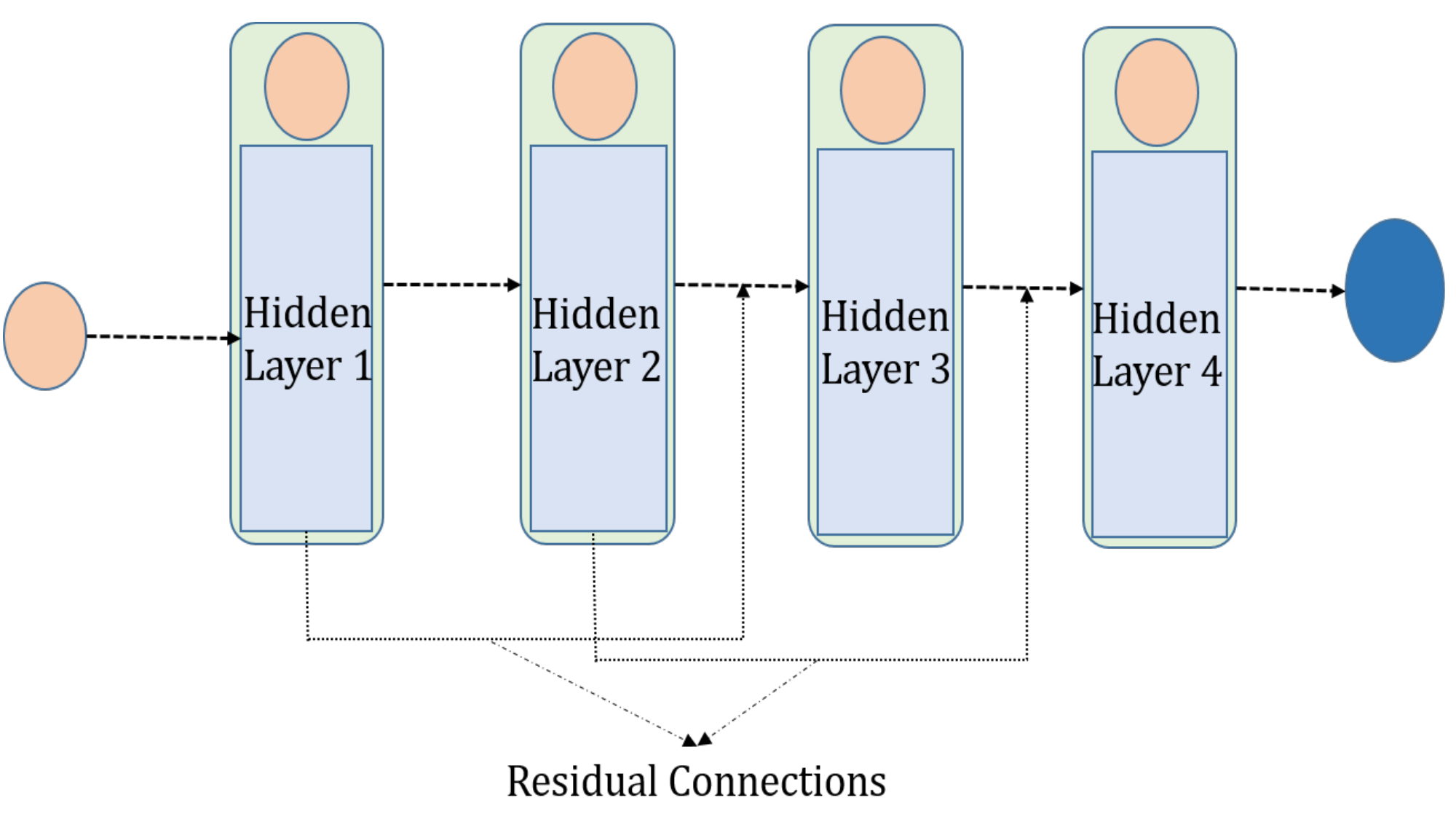}
    \caption{Physics informed residual learning framework for pricing European put options.}
    \label{pirl}
\end{figure}

With the development of PIDL (\cite{raissi}), the modelling of complex real-world phenomena through ML has gained popularity. Unlike the standard ML techniques, PIDL is a grey box modelling approach that utilises the representation power of deep learning (DL) along with the incorporation of theoretically proven mathematical models for generating a robust predictive model. One of the major milestones in the DL theory is the universal approximation theorem (\cite{hornik}) which states that a neural network with arbitrary depth and a fixed width can approximate any continuous function with desired precision. Based on this principle, the underlying framework for PIDL utilises the standard feed-forward neural network (FNN) which has proven its usefulness in various tasks like computer vision, finance, pattern recognition, time series analysis etc. (\cite{Alfaro2008, Angelini2008,chowdhary2020natural, fsi}). However, contradictory to the theoretical results, empirically it has been observed that the increase in the network depth can sometimes lead to the degradation problem which makes the optimisation of the loss functions difficult \cite{he2016deep}. To address this issue, a residual learning (RL) framework (\cite{he2016deep}), illustrated in figure \ref{pirl}, was introduced which establishes residual connections between the hidden layer that has been proven to mitigate some of the major problems present in the standard FNN framework. Moreover, the RL network shows improved signal propagation and smoother gradient flows. \cite{he2016deep, 8984747} thus allowing the training of deeper networks thereby enhancing the representation power of the DL even further. 

In general, an $L$-layer RL takes a $d$-dimensional vector as an input which is passed through $L-1$ intermediary layers known as the hidden layers. The $l^{\rm{th}}$ hidden layer transforms the input received from the $(l-1)^{\rm{th}}$ layer to $h^l$ dimensions through a linear or non-linear function applied upon an affine transformation. Finally, the $k$-dimensional output is generated in the $L^{\rm{th}}$ layer, known as the output layer, which is a sequential composition of the preceding hidden layers. Let $x_0\in\mathbb{R}^d$ be the input, then the RL can be formulated through the following set of equations

\begin{equation}
   f_l = \begin{cases} 
        \eta_l(W^{l}f_{l-1}+b^l), & \text{if } l = 1, L-1 \\
        \eta_l(W^{l}\bar{f}_{l-1}+b^l)+f_{l-1}, & \text{else} 
\end{cases}
\end{equation}
where $\eta_l$ is a linear or non-linear function known as the activation function applied upon the the $l^{\rm{th}}$ layer, $f_0 = x_0$, $f_l$ is the output generated in the $l^{\rm{th}}$ layer and $\bar{f}_l = \begin{pmatrix}f_l\\x_0\end{pmatrix}$.  $W^1\in\mathbb{R}^{n_1\times d},\;b^1\in \mathbb{R}^{n_1}$, $W^l\in\mathbb{R}^{n_l\times (n_{l-1}+d)},\;b^l\in\mathbb{R}^{n_l}$, $W^L\in\mathbb{R}^{k\times n_{L-1}},\;b\in\mathbb{R}^{k}$, are RL network parameters, where $n_l$ is the number of neurons in the $l^{th}$ layer.

The first phase of RL training requires the selection of a suitable architecture for representing the unknown solution of the considered system. In the second phase, the evaluation of the considered framework is accomplished through a suitable cost function details of which are provided in the next subsection.

\subsection{A description of physics embedding to the cost}
\label{PIC}
Consider a system defined by a set of partial differential equations of the form
\begin{equation}
    \begin{split}
        \partial_tV_i & = D_i[V_1, V_2, \Gamma],\; i=1,2, t\in[0,T];\\
        V_i(0, t) &= Ee^{-r(T-t)};\\
        V_i(S, T) &= \rm{max}(E-S, 0),
    \end{split}
\end{equation}
where $D_i,\;i=1,2$ are non-linear differential operators, $\Gamma\in\mathbb{R}^q$ is the vector of parameters of the considered model. Let the solution to the above system be approximated by the RL which is trained through randomly sampled set of points in the considered domain.

Let the state vector be denoted by $x_t\in\mathbb{R}^n$, for instance $x_t = S_t\in\mathbb{R}$ for Black-Scholes model and $x_t = (S_t, v_t)\in\mathbb{R}^2$ for the Heston stochastic volatility model, where $S_t$ and $v_t$ are the stock price and the volatility at time $t$ respectively. For generating the training data, we sample $N_\mathcal{A}$ realisations of $x_t$ in a closed convex set $\Omega\subseteq\mathbb{R}^n$ alongwith the model parameters, $\Gamma\in\mathbb{R}^q$, and $T$, the expiration date and $t(<T)$, the current time. Let $X_\mathcal{A} = (t^i, T^i, x^i_t, \Gamma^i)_{i=1}^{N_\mathcal{A}}$ be the sampled realisations of the current time, time to maturity, state variables and the model parameters. These set of points are known as the inner domain and will be employed for physics based regularisation. For measuring the error at the maturity of the contract, or the terminal boundary, generate $N_T$ realisations of the state variables, time to maturity and model parameters denoted by $X_T = (T^i, T^i, x^i, \Gamma^i)_{i=1}^{N_T}$. Similarly, $N_{low}$ realisations of the concerned variables are generated, denoted by $X_{low} = (t^i, T^i, x^i_t, \Gamma^i)$.

The RL network maps the above datasets to an approximate solution of the considered model by appropriately learning the trainable parameters $\Theta$ which is achieved by optimising a suitable cost function. Let $\bar{V} = (\bar{V}_1, \bar{V}_2)$ be the RL output obtained on the considered training data, then the cost function is defined as
\begin{equation}
    \mathcal{C}(\Theta) = \mathcal{C}_\mathcal{A}(\Theta)+\mathcal{C}_{T}(\Theta)+\mathcal{C}_{low}(\Theta).
\end{equation}
The first component of the cost, $\mathcal{C}_\mathcal{A}(\Theta)$, computes the error on the inner domain quantifying the accuracy of the RL at the time of expiry of the option and is given by
\begin{equation}
    \mathcal{C}_\mathcal{A}(\Theta) = \sum_{j=1}^{N_{\mathcal{A}}}\sum_{i=1}^2(\partial_t\bar{V}_i(X_\mathcal{A}^j)-D_i[\bar{V}_1, \bar{V}_2](X_\mathcal{A}^j))^2.
\end{equation}
The second component of the cost function, $\mathcal{C}_T(\Theta)$, computes the error between the RL output and the option payoff at expiry. It is defined as
\begin{equation}
    \mathcal{C}_T(\Theta) = \sum_{i=1}^{N_T}\sum_{i=1}^2(\bar{V}_i(X_T^j)-(H(S_T^j))^2.
\end{equation}
The third component of the cost function, $\mathcal{C}_{low}(\Theta)$, computes the error between RL output and discounted value of the option when the price of the underlying stock is at its minimum. It is defined as
\begin{equation}
    \mathcal{C}_{low}(\Theta) = \sum_{i=1}^{N_{low}}(\bar{V}_i(X_{low}^j)-e^{-r^j(T^j-t^j)}H(0))^2,
\end{equation}
where $r^j$, $T^j$ and $t^j$ are the risk free interest rate, time to maturity and the current time respectively, in the $j^{th}$ sample of the lower boundary sample set.

The cost function is optimised using numerical optimisation methods for updating the parameters of the PIRL network. In this work the limited memory Broyden–Fletcher–Goldfarb–Shanno or the L-BFGS-B (\cite{lbfgsb}) algorithhm is utilised for minimising the cost function. The PIRL network is implemented in python programming language using the Tensorflow 1.15 framework. The codes are available upon request.
\section{Physics informed residual learning framework: training and calibration procedure}
This section provides a detailed description of the dataset preparation and calibration procedure for training the PIRL network.

\subsection{Regime-switching Black-Scholes with PIRL network}
\begin{table}[h]
\centering
    \begin{tabular}{!{\vrule width 1.5pt}c|c|c|c!{\vrule width 1.5pt}}
    \noalign{\hrule height 1.5pt}
     Parameter & Range               & Parameter  & Range\\
    \noalign{\hrule height 1.5pt}
     $S_0$     &     $[40, 100]$     & $r$   & $[0.01, 0.025]$\\
     \hline
     $\sigma_1$     &     $[0.10, 0.30]$   & $\sigma_2$ & $[\sigma_1, 0.40]$\\
     \hline
     $T$ & [0, 4] & t & $[0, T]$\\
    \noalign{\hrule height 1.5pt}
    \end{tabular}
    \caption{Sampling ranges for the parameters for generating the training data for BSM regime-switching model.}
    \label{bspars}
\end{table}
\label{bspirl}
Since the BSM model utilises the geometric Brownian motion (GBM) for modelling the asset price, the state vector is $x_t = S_t$, i.e. the price of the asset at time $t$. Further, the parameters of the GBM under regime-switching are the risk free rate $r$ and the volatilities, $\sigma_1$ and $\sigma_2$, of the underlying asset under the two different economic states. Therefore, the input to the PIRL network for BSM model becomes $X = (t, T, S_t, r, \sigma_1, \sigma_2)$ where $t$ and $T$ are the current time and time of expiration, respectively, of the option hence, the auxilliary points, terminal boundary points and the lower boundary points, defined in \eqref{PIC}, respectively become $X_\mathcal{A} = (t^i, T^i, S_t^i, r^i, \sigma_1^i, \sigma_2^i)_{i = 1}^{N_\mathcal{A}}$, $X_T = (T^i, T^i, S_T^i, r^i, \sigma_1^i, \sigma_2^i)_{i=1}^{N_T}$, $X_{\rm{low}} = (t^i, T^i, 0, r^i, \sigma_1^i,\ \sigma_2^i)_{i=1}^{N_{\rm{low}}}.$ The cost of training is computed on these sets with $N_\mathcal{A} = 20000$, $N_T = 5000$ and $N_{\rm{low}} = 5000$. For simulation purposes the transition rates are assumed to be $\lambda_{12}=2\;\rm{and}\;\lambda_{21}=1$. For simulating a wide range of market conditions, the state variables and the parameters are randomly sampled from the ranges specified in Table \ref{bspars}. 

\subsection{Regime-switching Heston stochastic volatility with PIRL network}
\label{hestondata}
\begin{table}[h]
\centering
    \begin{tabular}{!{\vrule width 1.5pt}c|c|c|c!{\vrule width 1.5pt}}
    \noalign{\hrule height 1.5pt}
     Parameter & Range               & Parameter  & Range\\
    \noalign{\hrule height 1.5pt}
     $S_0$     &     $[40, 100]$     & $\rho$   & $[-0.85, -0.55]$\\
     \hline
     $v_0$     &     $[0.01, 0.1]$   & $\sigma_1$ & $[0.1, 0.45]$\\
     \hline
     $r$       &     $[0.015, 0.025]$& $\sigma_2$ & $[0.35, 0.75]$\\
     \hline
     $\kappa$  &     $[1.4, 2.6]$    & $T$        & $[0, 4]$\\
     \hline
     $\gamma$  &     $[0.01, 0.1]$   & $t$        & $[0, T]$\\ 
    \noalign{\hrule height 1.5pt}
    \end{tabular}
    \caption{Sampling ranges for the parameters for generating the training data for Heston stochastic volatility with regime switching model.}
    \label{hpars}
\end{table}
The Heston stochastic volatility model simulates the dynamics of an asset via a coupled system of stochastic differential equations governing the asset price and its volatility at time $t$. Therefore, the state vector is $x_t=(S_t, v_t)$ where $S_t$ and $v_t$ respectively, are the asset price and its volatility at time $t$. Furthermore, the model assumes that the asset price dynamics are governed by a fixed set of parameters namely the risk free rate, $r$, the long term mean, $\kappa$, the mean reversion rate, $\gamma$ and the volatilities of the volatility in the two different economic states, $\sigma_1$ and $\sigma_2$. Therefore, the PIRL network takes the state vector, the model parameters, current time and the time to maturity, i.e. $X_t = (t, T, (S_t, v_t), r, \kappa, \theta, \sigma_1, \sigma_2)$, as input and maps them to the European option price by utilising the principles of Heston model defined in \eqref{hestonpde}. Similar to the PIRL network for BSM, the auxilliary point, terminal boundary points and the lower boundary points respectively become, $X_\mathcal{A} = (t^i, T^i, (S_t^i, v_t^i), r^i, \kappa^i, \theta^i, \sigma_1^i, \sigma_2^i)_{i=1}^{N_\mathcal{A}}$, $X_T = (T^i, T^i, (S_T^i, v_T^i), r^i, \kappa^i, \theta^i, \sigma_1^i, \sigma_2^i)_{i=1}^{N_T}$ and $X_{\rm{low}}=(t^i, T^i, (0, v_t^i), ^i, \kappa^i, \theta^i, \sigma_1^i, \sigma_2^i)_{i=1}^{N_{\rm{low}}}$. The PIRL network computes the cost of training on these sets with $N_\mathcal{A}=30000,\;N_T=10000$ and $N_{\rm{low}}=10000.$ For the purpose of simulation the transition rates are assumed to be $\lambda_{12}=2$ and $\lambda_{21}=3$. Similar to the BSM described in section \ref{bspirl}. The input parameters are randomly sampled from a range specified in Table \ref{hpars}, such that a wide array of market conditions are covered.

\subsection{Calibrating procedure of the PIRL network}
In the two previous subsections, a brief description of the training data and the physics-informed cost function was provided. However, one of the biggest challenges in training all the DL models including the PIRL model is the selection of the most suitable, if not the optimal, architecture for representing the target output. This process for selecting the representative architecture for the DL framework is known as hyperparameter tuning. The hyperparameters are those parameters of the DL models that are not trainable via the optimisation algorithm and can be selected primarily through experimental observations. The important hyperparameters of the PIRL network include the network's depth, the width and the epochs.

For this work, the hyperparameters described above are selected by training the network with varying configurations. The networks are trained up to a fixed number of epochs for a fair comparison. The Tables \ref{bserrors} and \ref{hestonerrors} describe the errors obtained using different network configurations. Independent experiments were performed for the BSM and Heston stochastic volatility models to select the most suitable network architectures.

\begin{table}[h]
\centering
\begin{tabular}{!{\vrule width 1.5pt}c!{\vrule width 1.5pt}c!{\vrule width 1.5pt}c!{\vrule width 1.5pt}c!{\vrule width 1.5pt}c!{\vrule width 1.5pt}}
    \noalign{\hrule height 1.5pt}
 \textbf{Hidden Layers} & \textbf{Hidden Units} & \textbf{Physics Loss} & \textbf{Total Loss}& \textbf{Test Loss}\\
    \noalign{\hrule height 1.5pt}
  \multirow{3}{*}{4} & 16 & 0.0055 & 0.0069 & 0.0117\\
   \Cline{1.5pt}{2-5}

  & 32 & 0.0028 & 0.0031 & 0.0283\\
\Cline{1.5pt}{2-5}

  & 48 & 0.0007 & 0.0009 & 0.0013\\
\noalign{\hrule height 1.5pt}
  \multirow{3}{*}{6} & 16 & 0.0028 & 0.0035 & 0.0035\\
   \Cline{1.5pt}{2-5}
  & 32 & 0.0010 & 0.0012 & 0.0022\\
   \Cline{1.5pt}{2-5}
  & 48 & 0.0005 & 0.0006 & 0.0021\\
\noalign{\hrule height 1.5pt}
  \multirow{3}{*}{8} & 16 & 0.0013 & 0.0015 & 0.0006\\
    \Cline{1.5pt}{2-5}
  & 32 & 0.0008 & 0.0010 & 0.0046\\
   \Cline{1.5pt}{2-5}
  & 48 & 0.0005 &  0.0006 & 0.0010\\
\noalign{\hrule height 1.5pt}
\end{tabular}
\caption{PIRL architectures and corresponding losses in regime-switching BSM model.}
\label{bserrors}
\end{table}
\begin{table}[h]
\centering
\begin{tabular}{!{\vrule width 1.5pt}c!{\vrule width 1.5pt}c!{\vrule width 1.5pt}c!{\vrule width 1.5pt}c!{\vrule width 1.5pt}c!{\vrule width 1.5pt}}
    \noalign{\hrule height 1.5pt}
  \textbf{Hidden Layers} & \textbf{Hidden Units} & \textbf{Physics Loss} & \textbf{Total Loss} & \textbf{Test Loss}\\
    \noalign{\hrule height 1.5pt}
  \multirow{3}{*}{2} & 16 & 0.0821 & 0.1014 & 0.5541\\
   \Cline{1.5pt}{2-5}
    & 32 & 0.0371 & 0.0454 & 2.2771\\
\Cline{1.5pt}{2-5}
  & 48 & 0.0129 & 0.0162 & 0.1324\\
\noalign{\hrule height 1.5pt}
  \multirow{3}{*}{4} & 16 & 0.0177 & 0.0203 & 0.0430\\
    \Cline{1.5pt}{2-5}
  & 32 & 0.0050 & 0.0057 & 0.0105\\
   \Cline{1.5pt}{2-5}
  & 48 & 0.0047 & 0.0053 & 0.0464\\
\noalign{\hrule height 1.5pt}
  \multirow{3}{*}{6} & 16 & 0.0198 & 0.0229 & 0.2007\\
   \Cline{1.5pt}{2-5}
  & 32 & 0.0036 & 0.0039 & 0.0071\\
   \Cline{1.5pt}{2-5}
  & 48 & 0.0023 & 0.0025 & 0.2870\\
\noalign{\hrule height 1.5pt}
\end{tabular}
\caption{PIRL architectures and corresponding losses in regime-switching Heston stochastic volatility model.}
\label{hestonerrors}
\end{table}

\section{Numerical Results}
This subsection describes the numerical results obtained with the selected PIRL networks for BSM and Heston stochastic volatility with regime-switching cases. The results are generated on a fixed set of parameters of the models. To compare the option prices obtained by our methodology to the prices obtained from the alternative pricing techniques, we follow \cite{elliott2007pricing} and adopt the characteristic function approach for pricing options in the BSM regime switching model, whereas we adopt Monte Carlo simulations for the Heston's case. The details of these techniques are given in the Appendix A.

For quantifying the performance of the proposed pricing model we will use the following metrics
\begin{itemize}
    \item Mean squared error ($MSE$):
    $$MSE(V, \hat{V}) = \frac{1}{N}\sum_{i=1}^N(V_i-\hat{V}_i)^2.$$
    \item Mean absolute errror ($MAE$):
    $$MAE(V, \hat{V}) = \frac{1}{N}\sum_{i=1}^N|V_i-\hat{V}_i|.$$    
\end{itemize}

where $\hat{V} = \{\hat{V}_1,\hat{V}_2,\cdots,\hat{V}_N\}$ is the set of predictions for the ground truth $V = \{V_1,V_2,\cdots,V_N\}$.
\subsection{Black-Scholes with regime-switching}
Table \ref{bserrors} shows the total training loss and the physics loss for various architectures of PIRL network with the BSM regime-switching economy model and the corresponding test losses. The conducted experiments explore the efficiency of the PIRL by varying the number of hidden layers and the number of hidden units per layer. Further, the trained models are validated on 5000 test points, randomly sampled within the range described in the Table \ref{bspars}. The model with the least test loss is then selected for further analysis. The model selection is based on the test loss rather than the training loss, which alleviates the problems arising due to overfitting. It can be observed from the Table \ref{bserrors} that the 8 layers-deep model with 16 neurons per layer exhibits the least test loss in generating European option prices.

The selected PIRL network for the BSM regime-switching model after training was tested for generating the European option prices on a fixed set of parameters with $r = 0.02,\;\sigma_1 = 0.15,\;\sigma_2=0.35$. For validation, the obtained prices are compared with the analytical solution of the BSM regime-switching model described in the appendix.

Figure \ref{bsrsresults} illustrates the option prices obtained for the considered set of parameters with the underlying stock price varying from 30 to 110 keeping all the parameters fixed. Figure \ref{bsrsresults}(a) and (b) illustrate the European option prices at the maturity of the contract. It can be observed that the PIRL network perfectly captures the option prices at the maturity which is exactly equal to the payoff in both the regimes. Figure \ref{bsrsresults}(c) shows the price of the European options obtained with the analytical solution and the PIRL network for various stock prices 1 year before the maturity of the contract. The model efficiently captures the prices that precisely match the analytical option prices in both economic regimes. Furthermore, it can be observed from the plot that PIRL prices adhere to the condition $\underset{S\rightarrow\infty}{\lim}V_i(S, t)=0$. This observation is noteworthy as no prior information regarding this condition was supplied to the PIRL either in the form of physics loss or the training data. 

\begin{figure}[h]
 \begin{minipage}[b]{0.5\linewidth}
 \centering
     \includegraphics[width=0.9\linewidth, height=5cm]{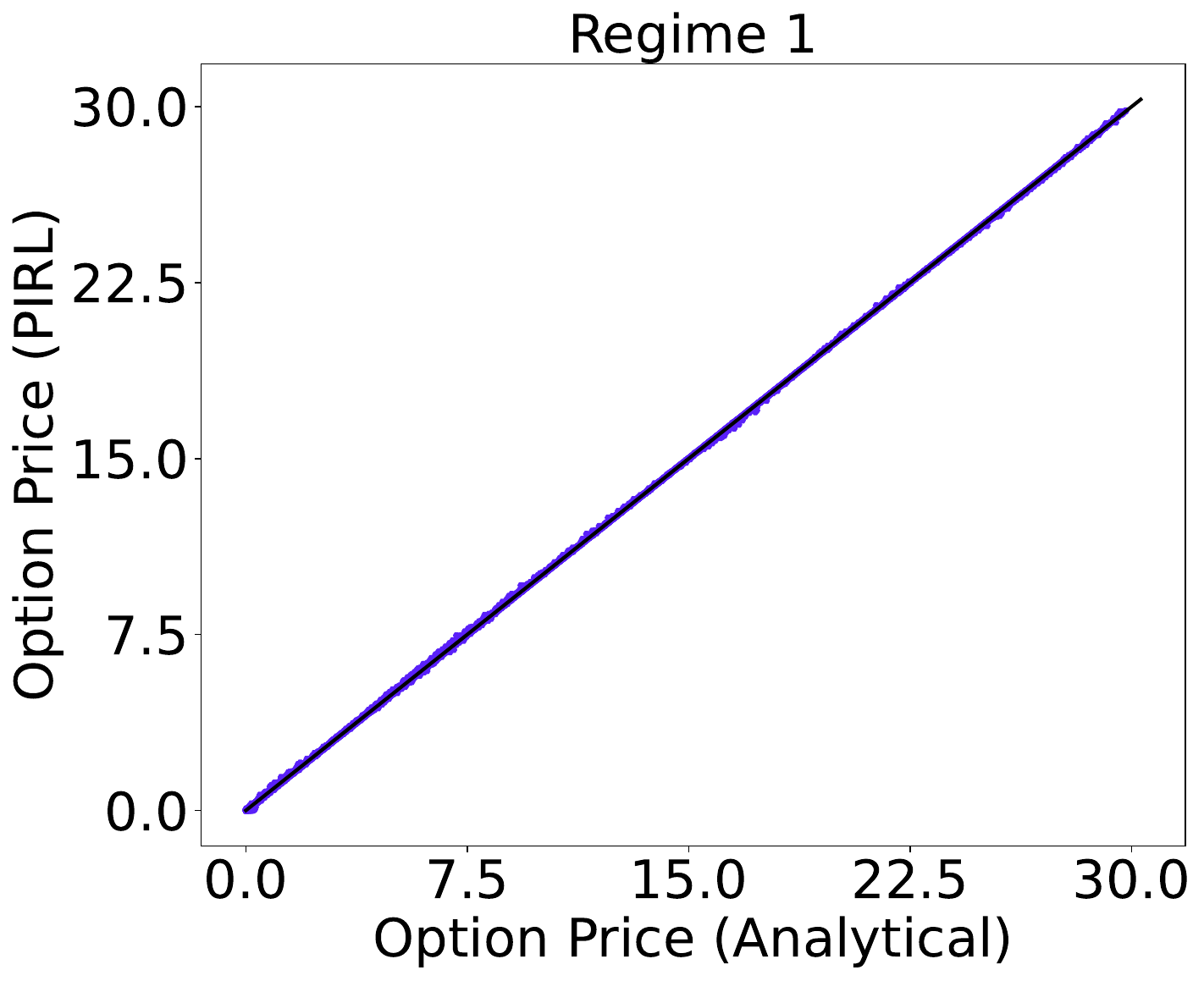}
     \caption*{(a)}
 \end{minipage}
 \begin{minipage}[b]{0.5\linewidth}
 \centering
     \includegraphics[width=0.9\linewidth, height=5cm]{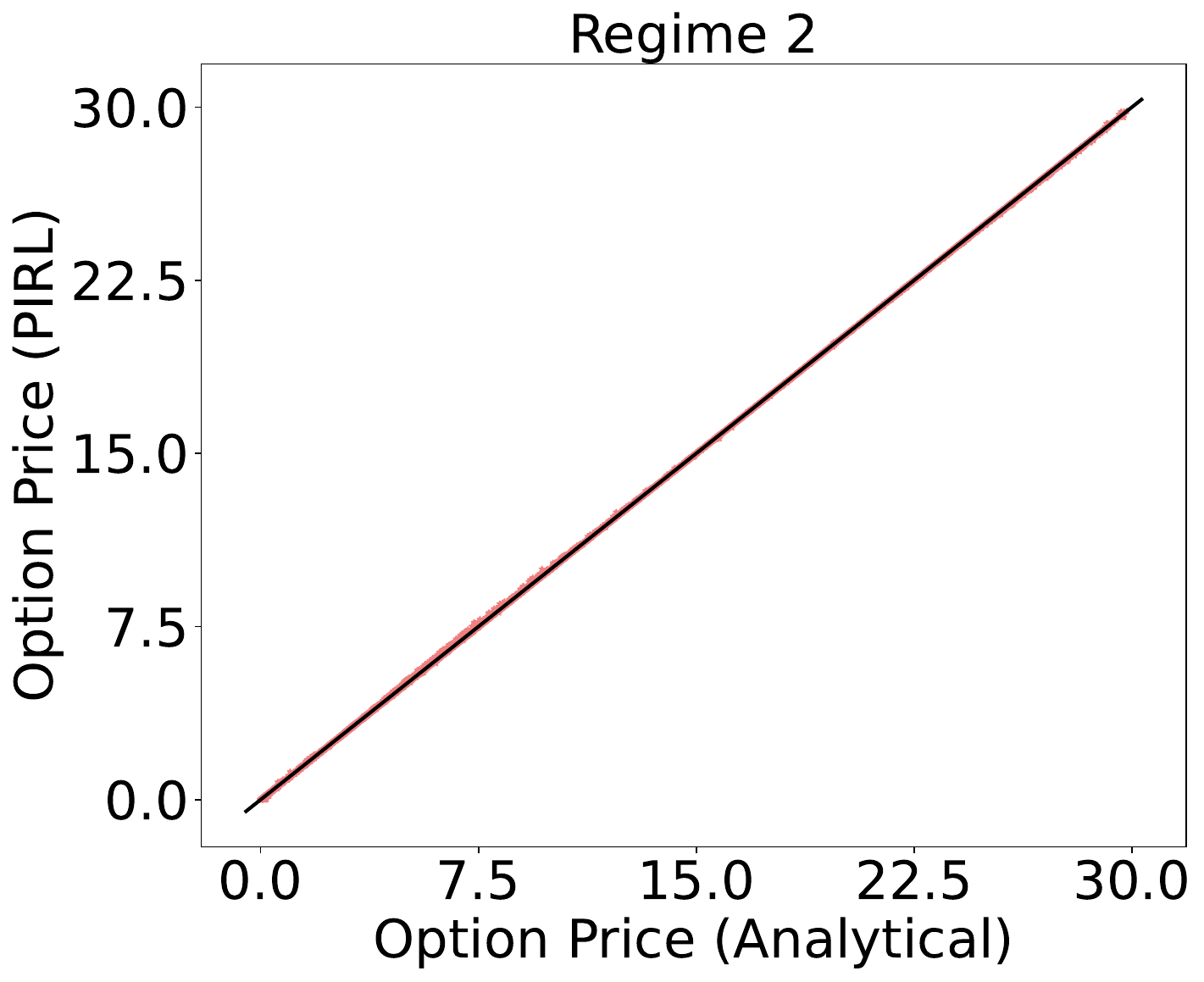}
     \caption*{(b)}
 \end{minipage}

 \caption{Put option prices in BSM regime-switching model obtained for 25000 different realisations of the state vector and model parameters via the analytic and PIRL pricing methods.}
 \label{bsrsresults2}
\end{figure}

For further analysis, 25000 realisations of the state vector and the model parameters along with the time to expiry were randomly sampled and the prices were calculated using both the analytical method and selected the PIRL network. The evaluation of the European option prices on this random sample with the analytical method took approximately 3 hours while the training of the PIRL network on the 30000 sample points described in the subsection \ref{bspirl} took approximately 18 minutes. However, after the PIRL model has been trained, the results of any sample can be instantly generated for any input set. Figure \ref{bsrsresults2} illustrates the European put option prices in the two different economic states obtained via the analytical solution and the PIRL network. The graph shows that most of the prices obtained lie in a straight line with very slight deviations. It is important to note that computing the prices analytically also requires numerical approximations thus allowing for some error in the true prices which explains the deviations of some prices from the straight line. Table \ref{errorsbs} report the considered error metrics for Figures \ref{bsrsresults} and \ref{bsrsresults2}. It can clearly be observed that the considered PIRL effectively captures the dynamics of the option prices governed by the BSM regime switching economy model both qualitatively and quantitatively.

\begin{figure}[h]
 \begin{minipage}[b]{0.3\linewidth}
 \centering
     \includegraphics[width=5cm, height=4cm]{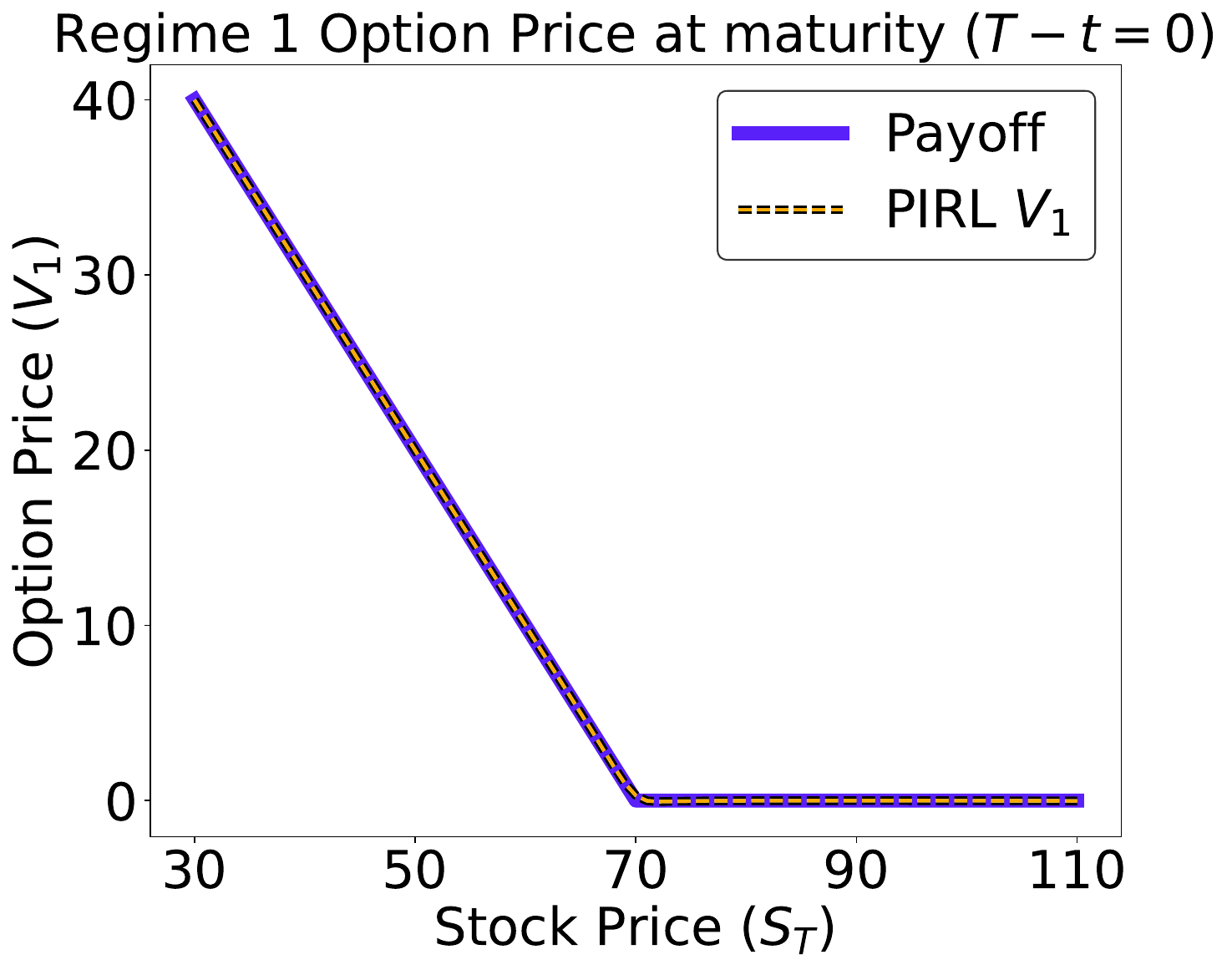}
     \caption*{(a)}
 \end{minipage}
 \begin{minipage}[b]{0.3\linewidth}
 \centering
     \includegraphics[width=5cm, height=4cm]{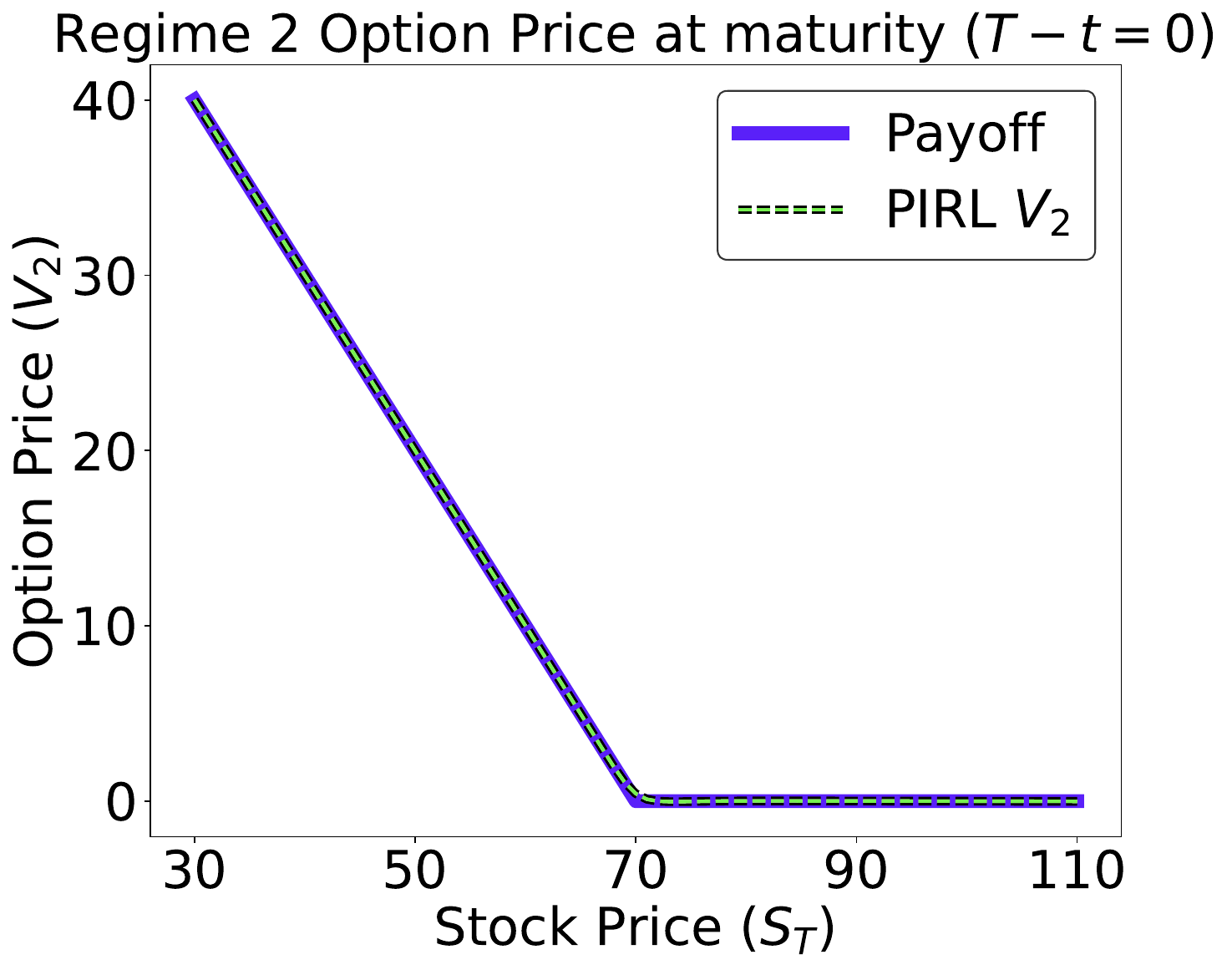}
     \caption*{(b)}
 \end{minipage}
 \begin{minipage}[b]{0.3\linewidth}
 \centering
     \includegraphics[width=5cm, height=4cm]{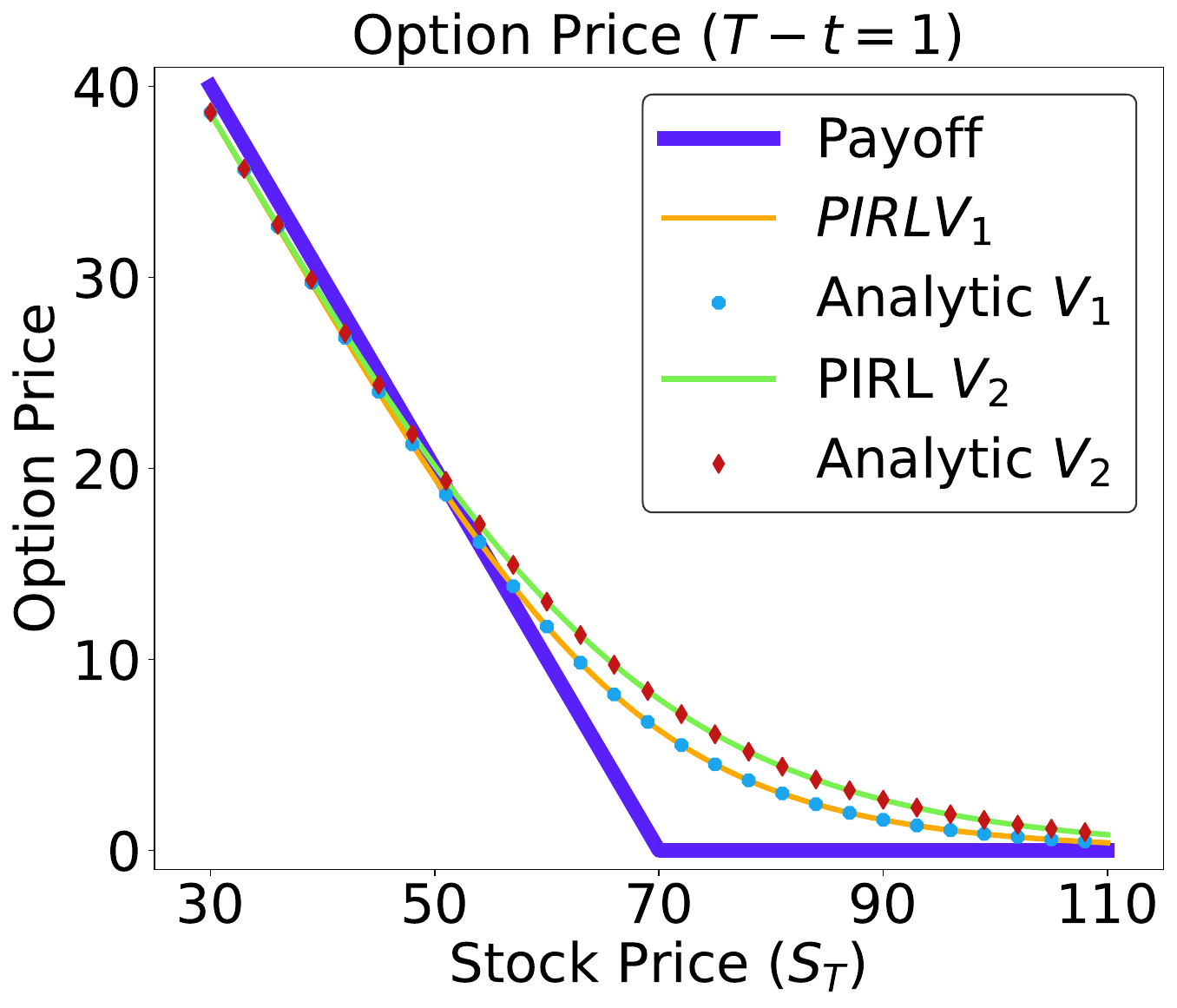}
     \caption*{(c)}
 \end{minipage}
 \caption{1 year European put options with strike price 70 in different economic states priced with PIRL network and the analytic method of regime-switching BSM model.}
\label{bsrsresults}
\end{figure}

\begin{table}[h!]
    \centering
    \setlength{\extrarowheight}{3pt} 
    \begin{tabular}{!{\vrule width 2pt}c!{\vrule width 2pt}c|c!{\vrule width 2pt}c|c!{\vrule width 2pt}}
        \noalign{\hrule height 2pt}
    \multirow{2}{*}{\textbf{Figure}} &  \multicolumn{2}{c!{\vrule width 2pt}}{\textbf{Regime 1}} & \multicolumn{2}{c!{\vrule width 2pt}}{\textbf{Regime 2}} \\
    \cline{2-5}
    & \textbf{MAE} & \textbf{MSE}  & \textbf{MAE} & \textbf{MSE} \\
    \noalign{\hrule height 2pt}
        \multicolumn{1}{!{\vrule width 2pt}c!{\vrule width 2pt}}{Figure \ref{bsrsresults} (a) and (b)}& 0.0092 & 0.0005  & 0.0092 & 0.0012 \\
    \cline{1-5}
    \multicolumn{1}{!{\vrule width 2pt}c!{\vrule width 2pt}}{Figure \ref{bsrsresults} (c)} & 0.0152 & 0.0006  & 0.0225 & 0.0013 \\
    \hline
    \multicolumn{1}{!{\vrule width 2pt}c!{\vrule width 2pt}}{Figure \ref{bsrsresults2}} & 0.0154 & 0.0009 & 0.0148 & 0.0008\\
    \noalign{\hrule height 2pt}
    \end{tabular}
    \caption{Out of sample prediction errors obtained for the BSM with regime switching economy model.}
    \label{errorsbs}
\end{table}

\subsection{Heston stochastic volatility with regime-switching}
The results reported in Table \ref{hestonerrors} describe the training and test losses obtained with the PIRL model equipped with the Heston stochastic volatility model with regime-switching economy. Similar to the case of the BSM, various configurations were trained with the training data described in section \ref{hestondata}. Further, the trained models were tested with testing data generated within the range outlined in Table \ref{hpars}. As in the previous subsection, the framework exhibiting the least test loss is selected for further investigation. From the Table \ref{hestonerrors} it can be concluded that the 6-layer deep network with 32 neurons per layer is the most suitable architecture for further investigations.

Further investigations with the selected architecture are then performed by generating European put prices and the results are compared with the Monte Carlo simulations. The results are generated by fixing the parameters with $v_0 = 0.05,\;=0.02,\;\kappa=2,\;\gamma=0.1,\;\sigma_1=0.25,\;\sigma_2=0.5,\;\rho=-0.8$. Since Monte Carlo simulations are expensive we restrict ourselves to the cases where we fix all the parameters except for time to maturity.

\begin{figure}[h]
    \centering
    \begin{tabular}{ccc}
         \includegraphics[width=5cm, height=4cm]{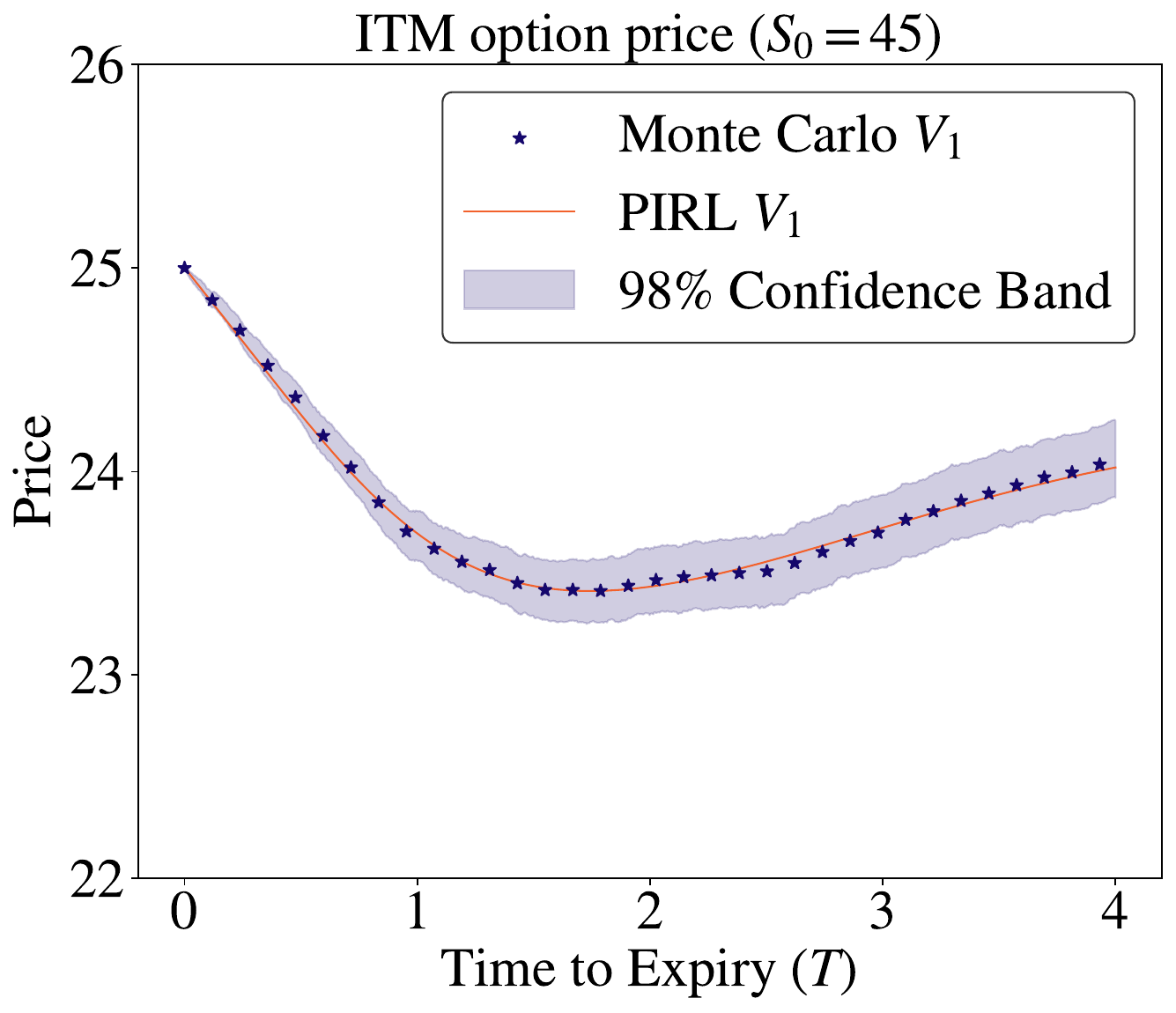} &
         \includegraphics[width=5cm, height=4cm]{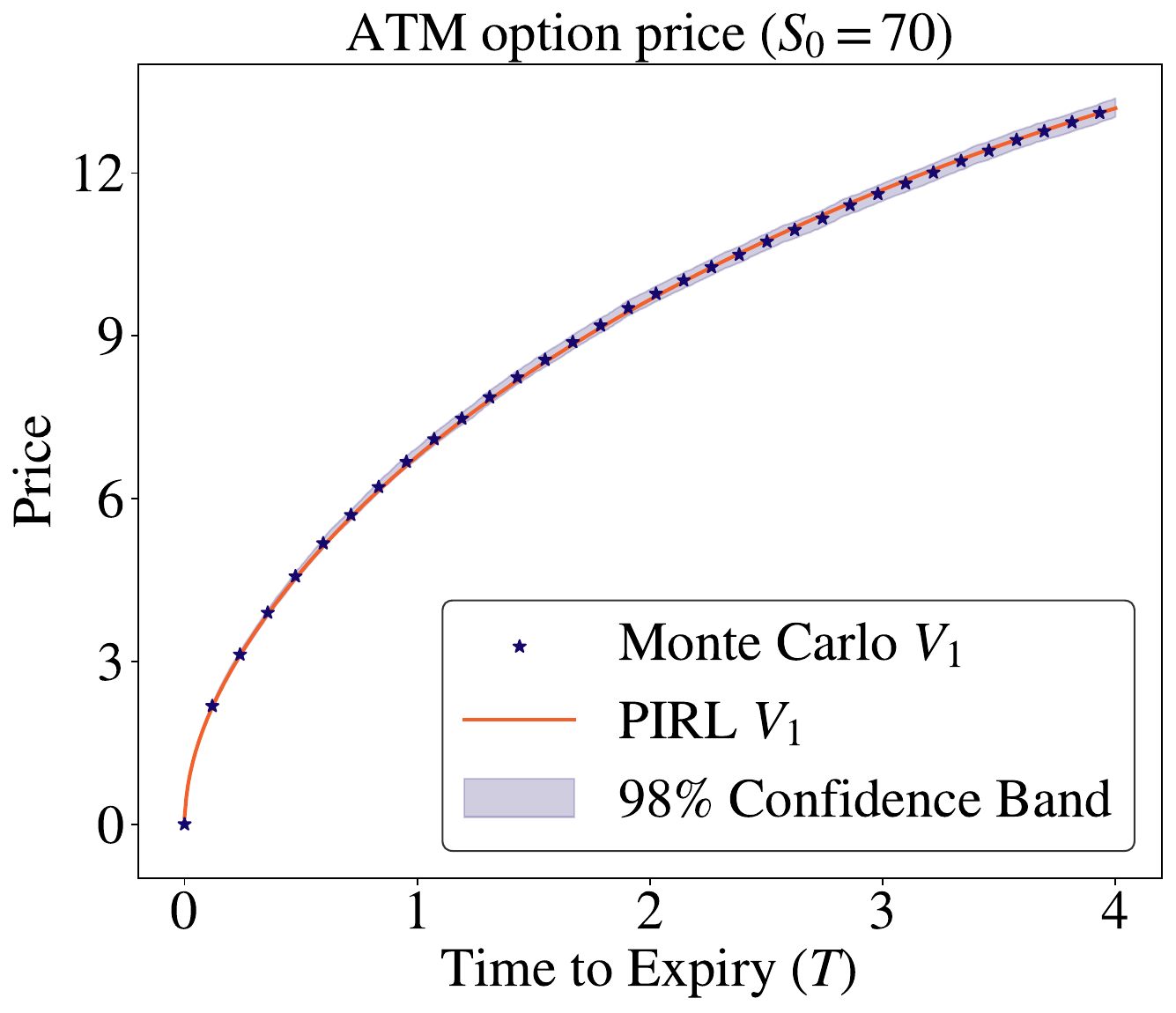}&
         \includegraphics[width=5cm, height=4cm]{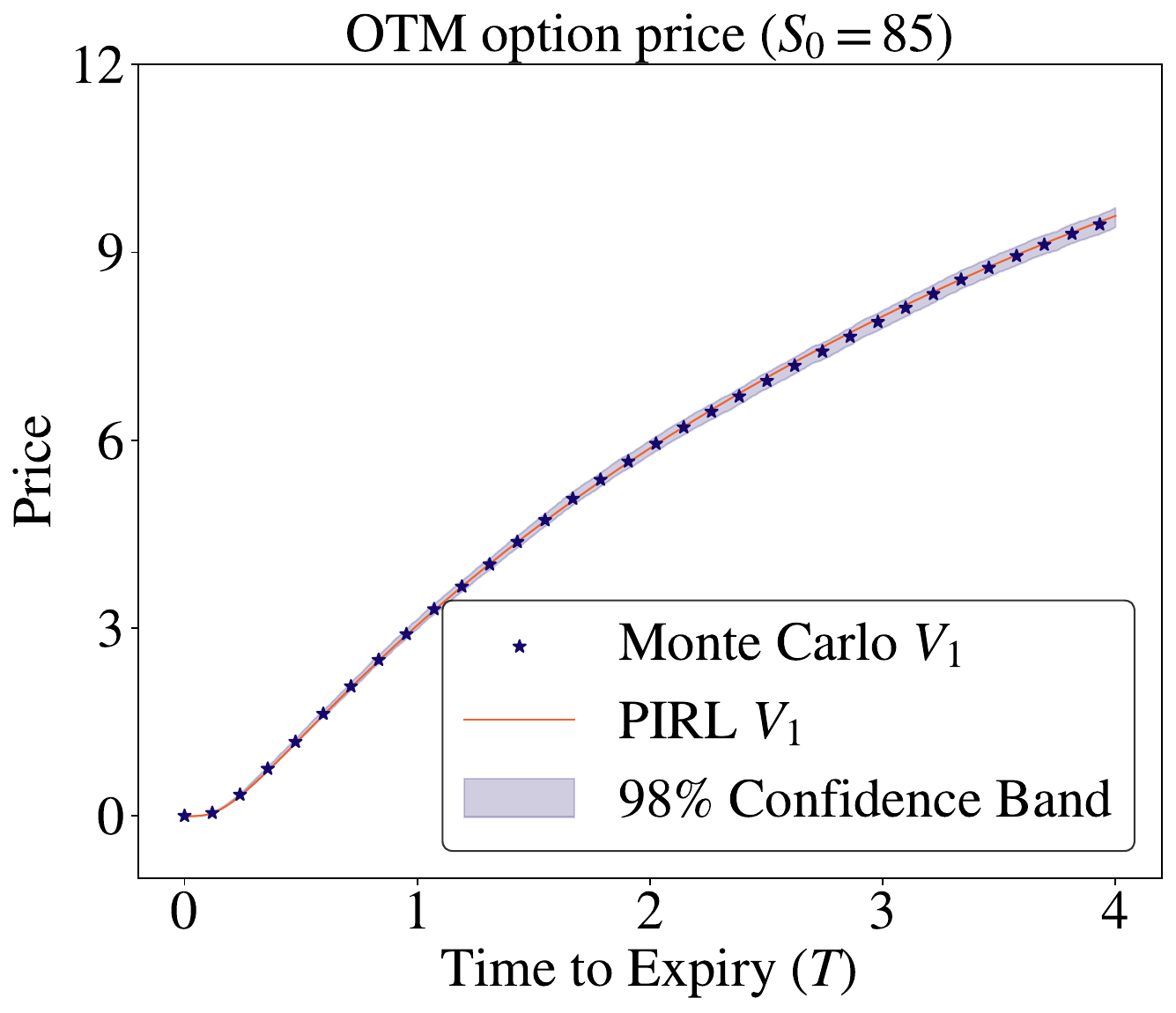}\\
         (a) & (b) & (c)\\
         \includegraphics[width=5cm, height=4cm]{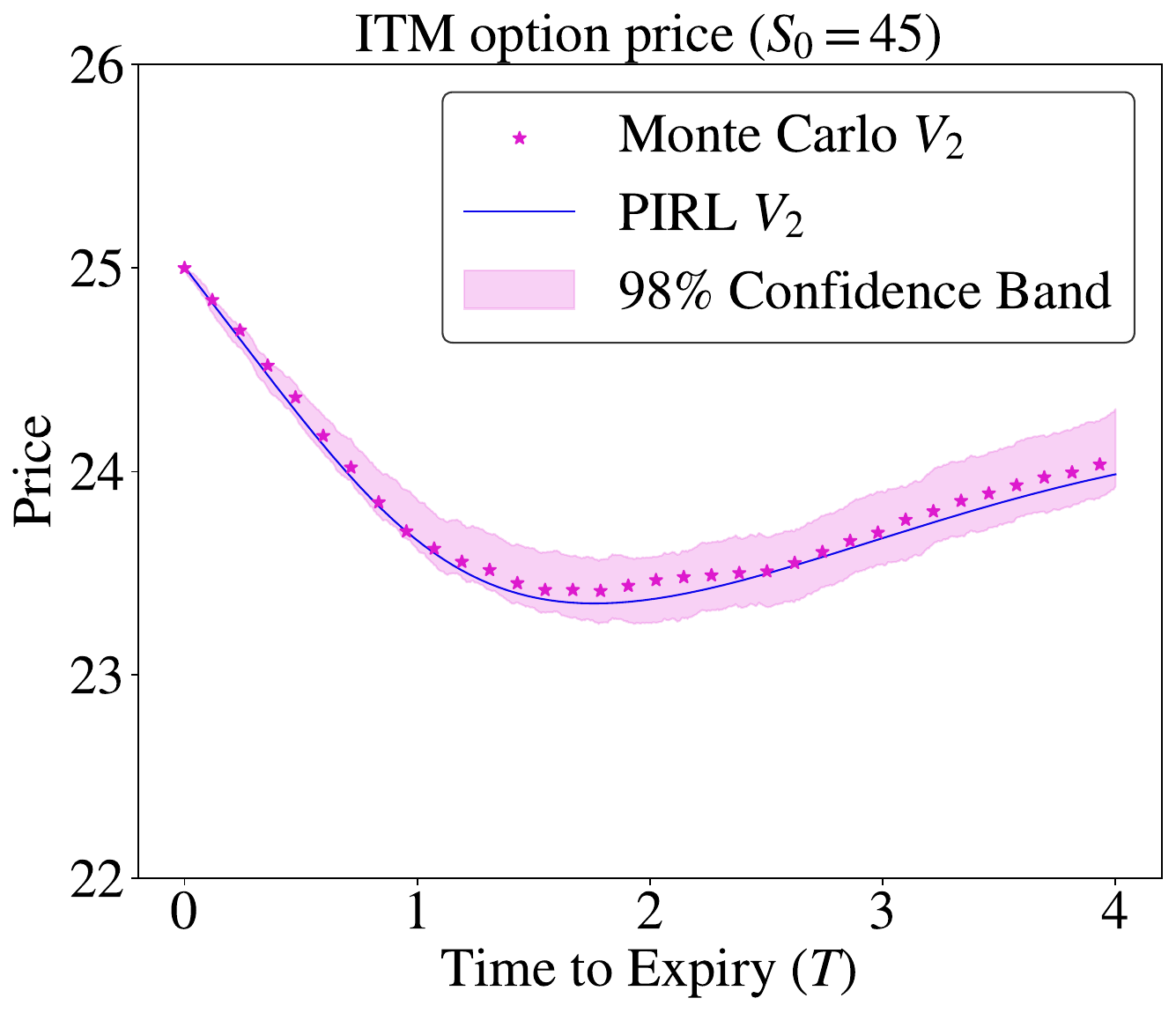}&
         \includegraphics[width=5cm, height=4cm]{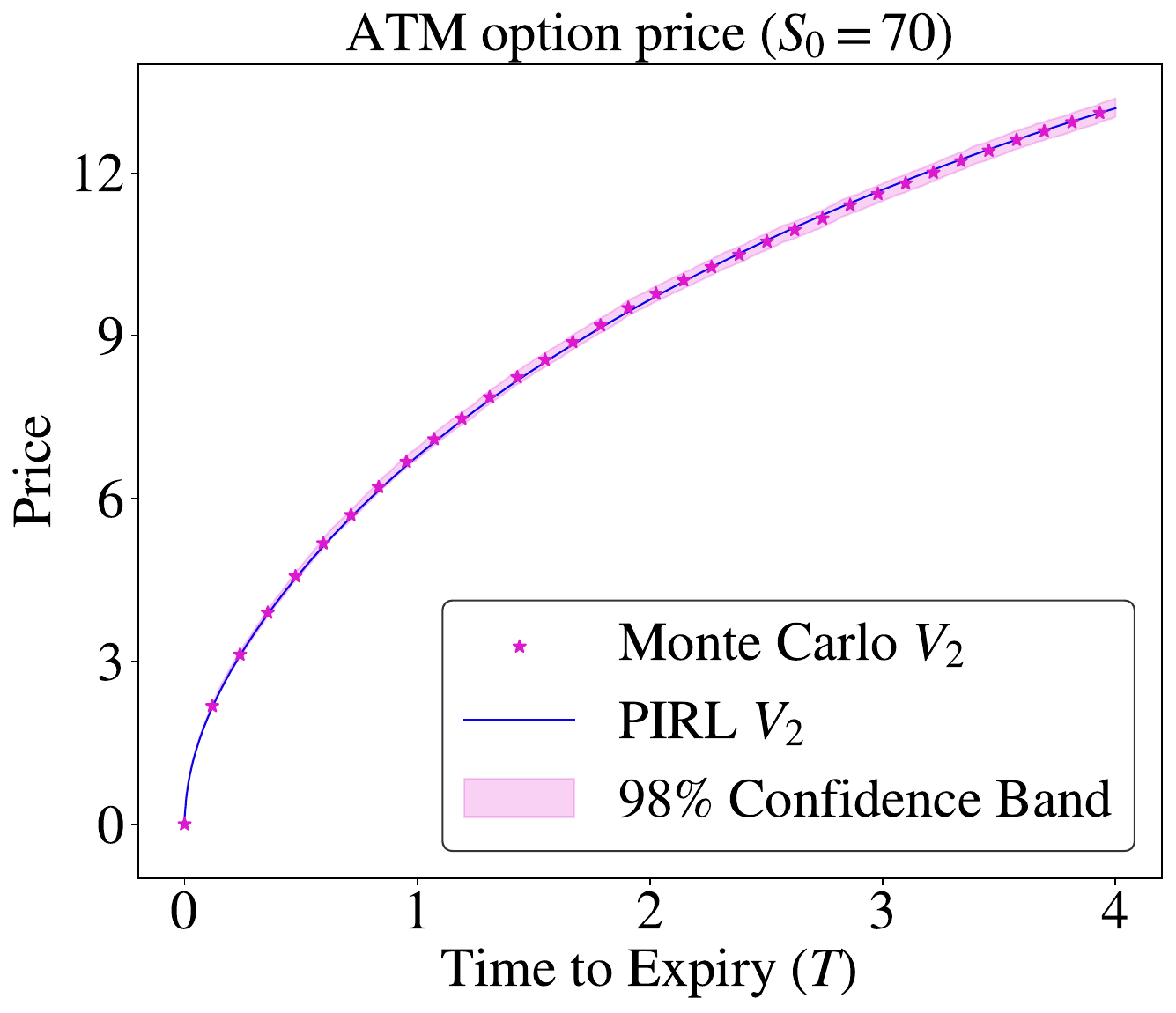}&
         \includegraphics[width=5cm, height=4cm]{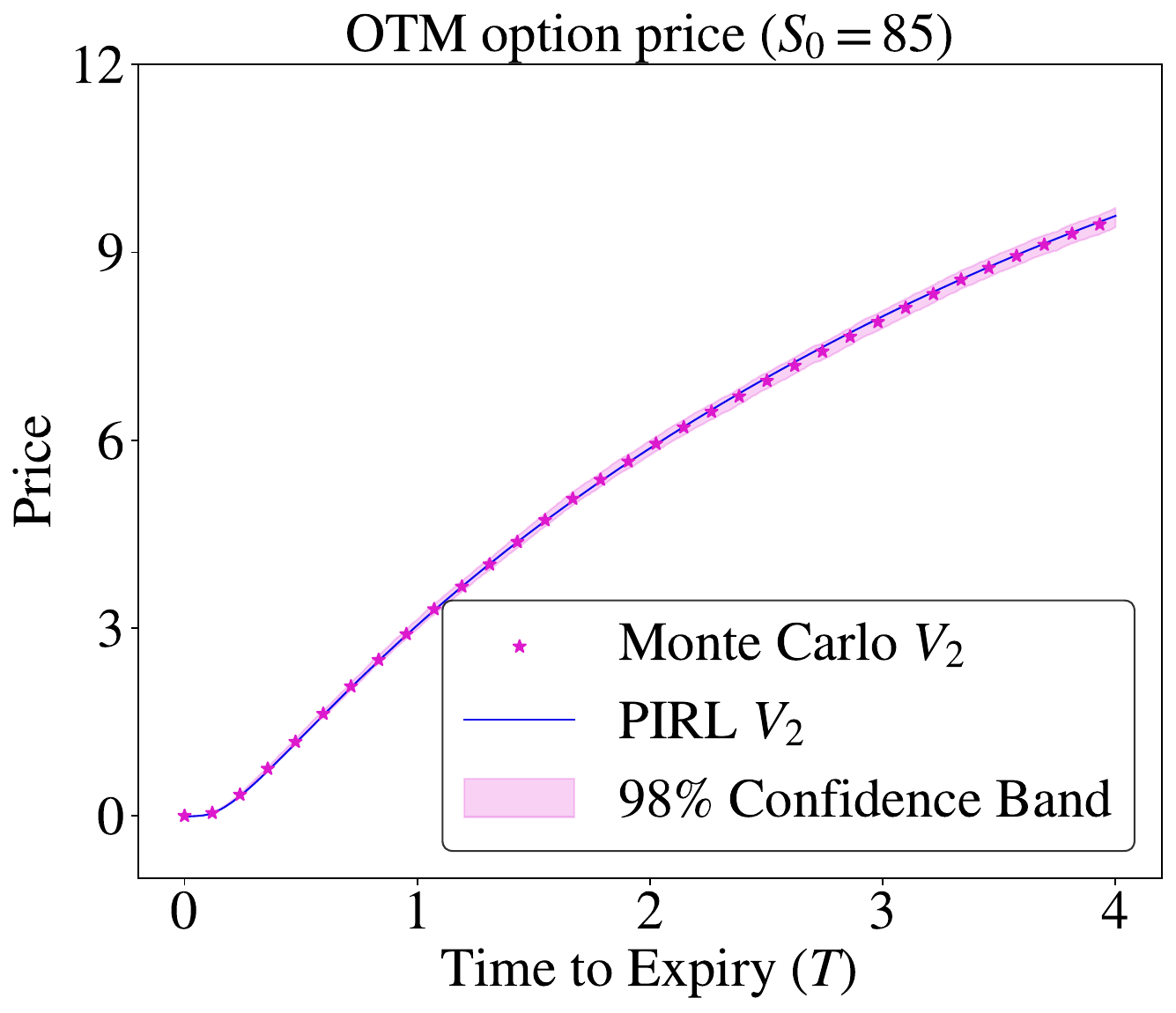}\\
         (e) & (f) & (g)
    \end{tabular}
    \caption{European put option prices in the two regimes under Heston stochastic volatility model with $\sigma_1=\sigma_2=0.4.$}
    \label{hestonrs}
\end{figure}

Figure \ref{hestonrs} demonstrates the European option prices by varying the time to maturity of the option  from 0 to 4 years, obtained through the PIRL network and the Monte Carlo simulations. The results are presented for in-the-money (ITM), at-the-money (ATM) and out- of-the-money (OTM) cases. The observations indicate that the model efficiently captures the put prices in both regimes and the predictions for the put options lie well within the 98\% confidence interval of the Monte Carlo prices. The most interesting observations arise in the case of the ITM option prices. The PIRL network well captures the initial dip in the option prices for approximately the first two years which increases afterwards for the next two years.Table \ref{errorshrs} report the MAE and MSE obtained for the considered instances of option prices. The error analysis indicates that the model is able to model the put prices with a reasonably good precision.

\begin{figure}
    \centering
    \begin{tabular}{ccc}
        \includegraphics[width=5cm, height=4cm]{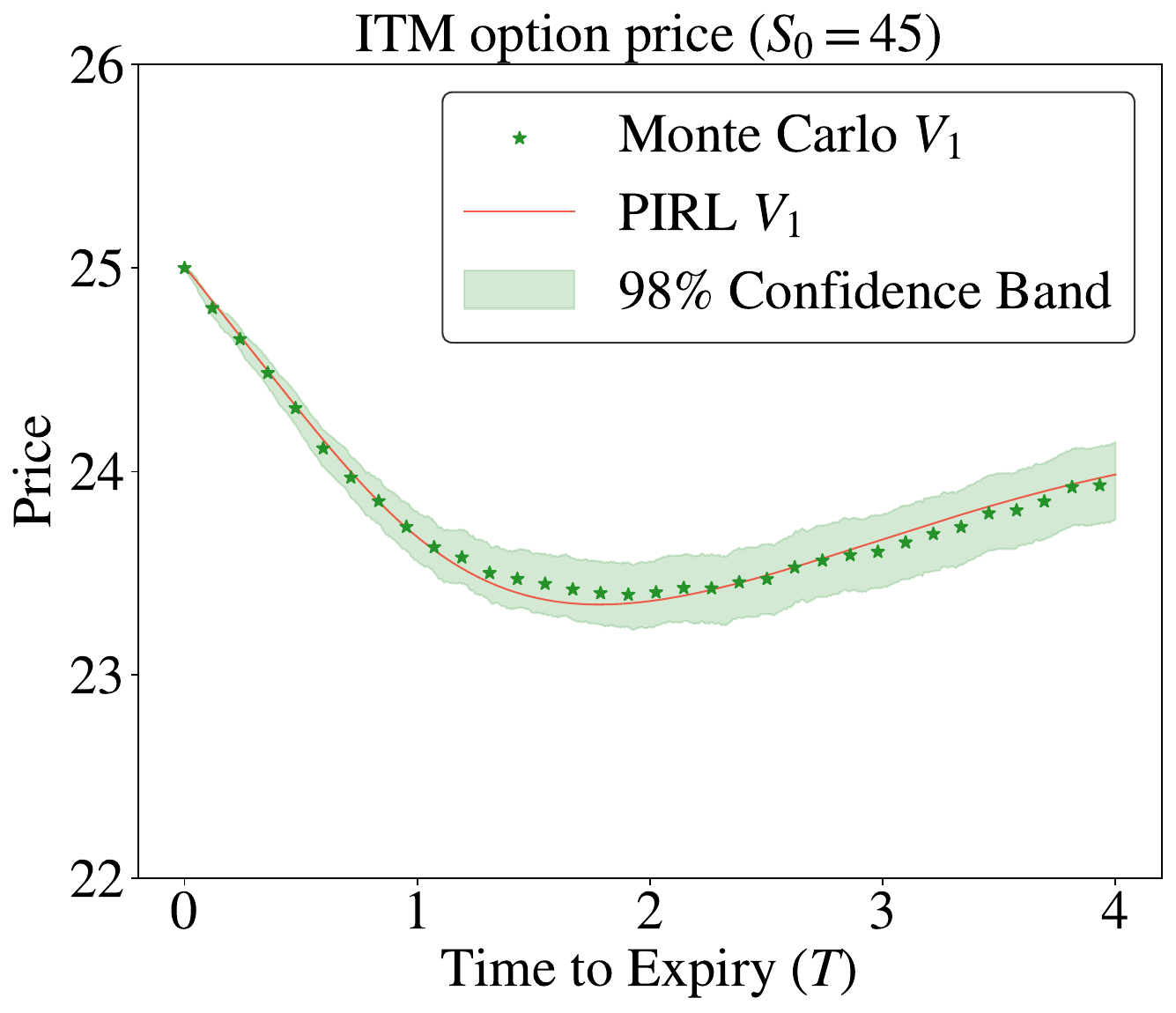} &
        \includegraphics[width=5cm, height=4cm]{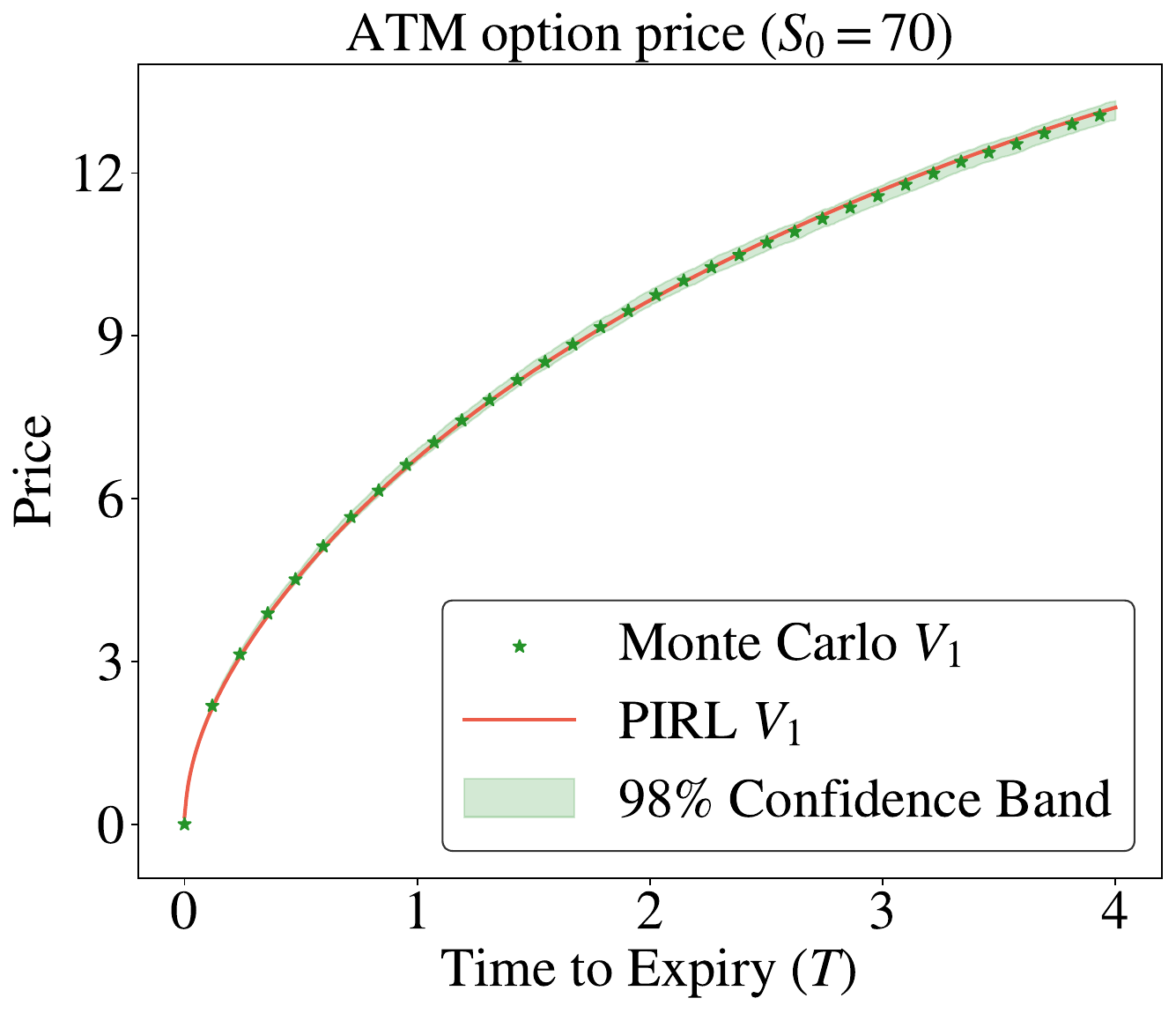}&
        \includegraphics[width=5cm, height=4cm]{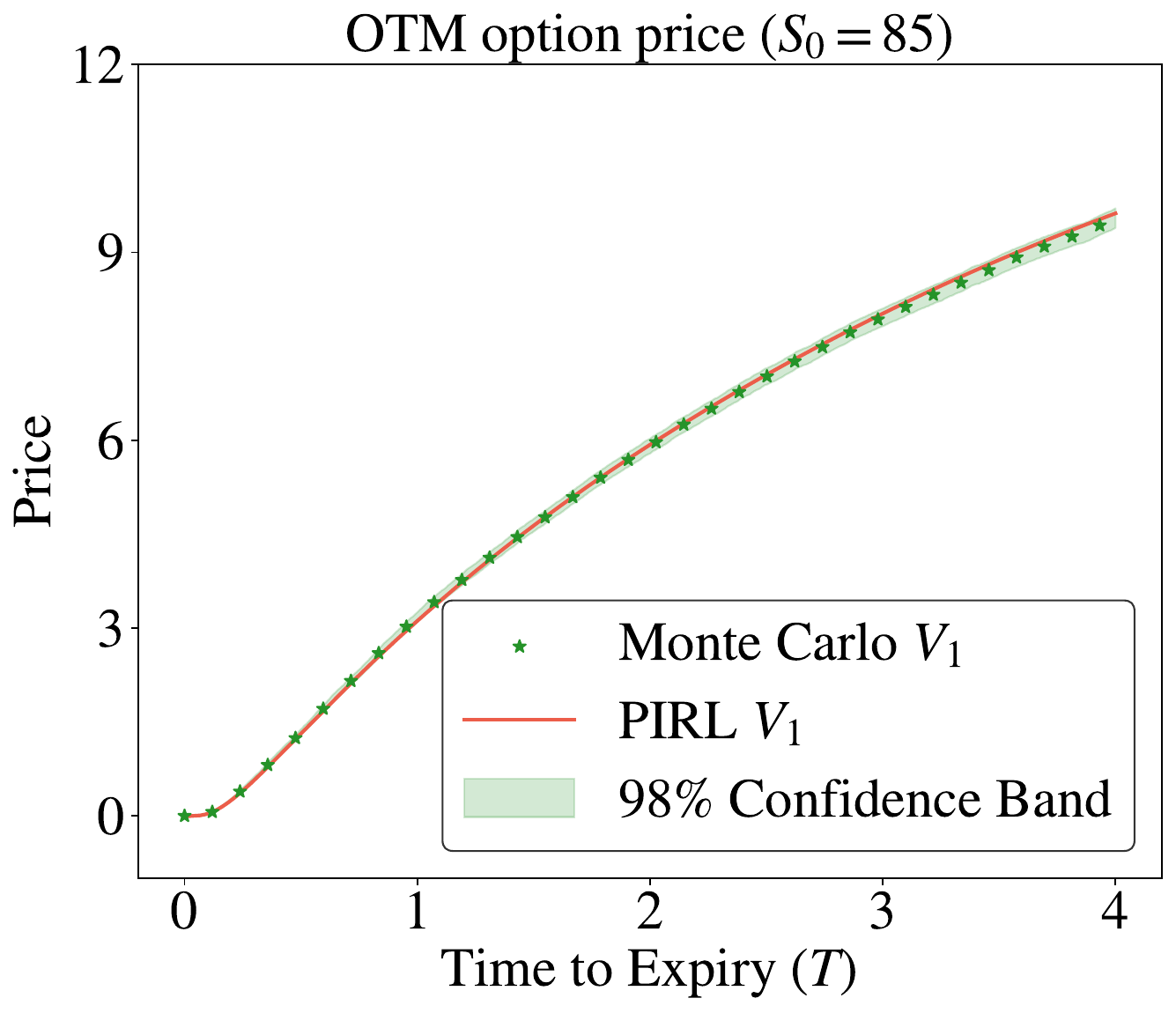}\\
        (a) & (b) & (c)
    \end{tabular}
    \caption{Heston option prices obtained through the Monte Carlo method and the Heston PIRL model with $\sigma_1=\sigma_2=0.4$.}
    \label{hestonstd}
\end{figure}

Note that, the overlap in the ranges of the volatilities of the volatility, i.e. $\sigma_1$ and $\sigma_2$, allows the PIRL to model the standard Heston stochastic volatility model without regime switching. This can be achieved by setting $\sigma_1=\sigma_2$ for both the regimes. The Figure \ref{hestonstd} displays the results generated with PIRL for the Heston stochastic volatility case. The results show that the model effectively aligns with the Monte Carlo simulation prices for all the ITM, ATM and OTM cases. Table \ref{errorsh} shows the MAE and MSE obtained for the considered instances of put option prices, which describes the efficacy of the PIRL network in learning the Heston option dynamics without being explicitly trained with PDE governing the Heston stochastic volatility model dynamics. All the obtained results suggest that the residual learning framework alongwith physics informed regularisation efficiently models the option price dynamics both quantitatively as well as qualitatively.

\begin{table}[h!]
    \centering
    \setlength{\extrarowheight}{3pt} 
    \begin{tabular}{!{\vrule width 2pt}c!{\vrule width 2pt}c|c!{\vrule width 2pt}c|c!{\vrule width 2pt}}
    \noalign{\hrule height 2pt}
    \multirow{2}{*}{\textbf{Option Type}} &  \multicolumn{2}{c!{\vrule width 2pt}}{\textbf{Regime 1}} & \multicolumn{2}{c!{\vrule width 2pt}}{\textbf{Regime 2}} \\
    \cline{2-5}
    & \textbf{MAE} & \textbf{MSE}  & \textbf{MAE} & \textbf{MSE} \\
    \noalign{\hrule height 2pt}
    ITM & 0.0199 & 0.0005  & 0.0589 & 0.0043\\
    \cline{1-5}
    ATM & 0.0345 & 0.0015  & 0.0357 & 0.0017 \\
    \cline{1-5}
    OTM & 0.0255 & 0.0009  & 0.0960 & 0.0135 \\
    \noalign{\hrule height 2pt}
    \end{tabular}
    \caption{Out of sample prediction errors for the Heston stochastic volatility with regime switching put option prices for ITM, ATM and OTM option types.}
    \label{errorshrs}
\end{table}
\begin{table}[h!]
    \centering
    \setlength{\extrarowheight}{3pt} 
    \begin{tabular}{!{\vrule width 2pt}c!{\vrule width 2pt}c|c!{\vrule width 2pt}}
    \noalign{\hrule height 2pt}
    \multirow{2}{*}{\textbf{Option Type}} &  \multicolumn{2}{c!{\vrule width 2pt}}{\textbf{Heston Price}}\\
    \cline{2-3}
    & \textbf{MAE} & \textbf{MSE}\\
    \noalign{\hrule height 2pt}
    ITM & 0.0337 & 0.0015\\
    \cline{1-3}
    ATM & 0.0442 & 0.0024\\
    \cline{1-3}
    ATM & 0.0436 & 0.0028\\
    \noalign{\hrule height 2pt}
    \end{tabular}
    \caption{Out of sample prediction errors for the Heston stochastic volatility option prices for ITM, ATM and OTM option types.}
    \label{errorsh}
\end{table}
\section{Conclusion}
In this work, two option pricing frameworks are presented utilising the PIRL by augmenting the physical laws of the BSM model and the Heston stochastic volatility model with regime-switching economic states. Furthermore, the proposed models take advantage of the residual learning framework for pricing the European put options, thus alleviating the limitations of the standard DL architecture by incorporating residual connections between the hidden layers. This allows training of deeper networks thus increasing the representation power of the standard FNN architecture.
Furthermore, integration of the considered mathematical models of option pricing leads to an increase in the predictive capabilities of the residual learning framework. This work aims to provide an alternative to standard option pricing methods.

The primary objective is to investigate the capabilities of the PIRL method in evaluating the European put option prices when the various parameters affecting the stock prices switch their states governed by a continuous-time Markov chain. For analysing the regime-switching properties of the options through the lens of PIRL, the BSM and the Heston stochastic volatility models were selected. Furthermore, an analytical expression for evaluating the European option prices for the BSM model was also presented which is used for validation of PIRL results. For the Heston model, the Monte Carlo method for pricing the European put options was used for the validation of corresponding the PIRL results.

Several experiments were performed for the selection of the most suitable architecture of the PIRL for both the BSM and Heston model cases. Furthermore, a wide range of market conditions were simulated to investigate the capabilities of the PIRL. The analysis showcased the promising results generated by the PIRL in valuing the put option prices in both the considered cases. The model efficiently captured the dynamics of the option prices in the cases of ITM, ATM and OTM option types. Furthermore, the model is able to generalise the results for the standard Heston stochastic model, i.e. without regime-switching volatility of the volatility. The PIRL results shows that the model effectively captured the intricate dynamics of option prices of all the considered mathematical models.

The presented results indicate that the main advantage of the PIRL lies in fast and efficient computation of the option prices for a wide array of market parameters thereby enabling the traders to execute timely trades or generate short-term strategies. The traditional methods are sensitive to model parameters and require reevaluation even with a small change in model parameters, however, the PIRL does not suffer from this drawback and can instantly generate the results for a large variety of market parameters after training.

Even though the PIRL performed exceptionally well and generated the desired results, it requires calibration and selection of a suitable architecture which best represents the target solution. No proper method does not yet exist for such a task and can only be performed using experimental observations. The proposed model well captures the dynamics of the considered pricing models, however, one may also explore the more complex option pricing model for instance jump diffusive models or fractional Brownian motion etc. for pricing the options with PIRL. Furthermore, one may also utilise a different DL architecture like LSTM or transformers etc. Since the PIRL shows exceptional capabilities in effectively pricing the European put prices, one may also explore its capabilities in modelling path dependent options like Bermudan options, Asian options or barrier options etc.

\section*{Acknowledgement}
This work was partially supported by the FIST program of the Department of Science and Technology, Government of India, Reference No. SR/FST/MS-I/2018/22(C).

\bibliographystyle{apalike}
\bibliography{bib.bib}
\newpage
\appendix
\section{\textbf{Appendix A}}
\subsection*{\textbf{Regime switching Black-Scholes model}}
\label{bsrsapp}
Under the risk-neutral martingale measure $Q$, the price of a European put option with maturity $T$ and strike price $E$, is given by
	\begin{eqnarray}
		\nonumber P&=&\mathbb{E}_t^Q\bigg(e^{-r(T-t)}(E-S_T)^+\bigg)\\
		\nonumber &=&e^{-r(T-t)}\mathbb{E}_t^Q\bigg(EI_{\{S_T\leq E\}}\bigg)-e^{-r(T-t)}\mathbb{E}_t^Q\bigg(S_TI_{\{S_T\leq E\}}\bigg)\\
  \label{BSPut}&=& e^{-r(T-t)}E P(S_T\leq E )-e^{-r(T-t)}\mathbb{E}_t^Q\bigg(S_TI_{\{S_T\leq E\}}\bigg),
	\end{eqnarray}
where $I_{A}$ is the indicator function of the set $A$. The first term, $P(S_T\leq E )$ is given by (\cite{shephard1991characteristic}),
\begin{equation*}
\mathbb{P}(S_T\leq E)=\frac{1}{2}-\frac{1}{\pi}\int_0^{+\infty}\rm{Real}\bigg(\frac{e^{-\hat{i}\eta\log(K)}f_1(\eta,t)}{j\eta}\bigg)d\eta,
\end{equation*}
with $\hat{i}=\sqrt{-1}$ and $f_1(\eta,t)=\mathbb{E}_t^{Q}\big(e^{\hat{i}\eta Y_T}\big)$ is the characteristic function of the $\log$ price, i.e., $Y_T=\log(S_T)$. The second term $\mathbb{E}_t^Q\bigg(S_TI_{\{S_T\leq E\}}\bigg)$ in Equation (\ref{BSPut}) can be simplified using $S_t$ as a numeraire as follows
\begin{equation*}
\frac{dQ_1}{dQ}=e^{-r(T-t)}\frac{S_T}{S_t},    
\end{equation*}
where $Q_1$ is a new measure under which we have 
\begin{equation*}
\mathbb{E}_t^Q\bigg(S_TI_{\{S_T\leq E\}}\bigg)=\mathbb{E}_t^{Q_1}\big(I_{\{S_T\leq E\}}\big)=\tilde{\mathbb{P}}(S_T\leq E),
\end{equation*}
with $\tilde{\mathbb{P}}$ denoting the probability under measure $Q_1$. Following \cite{shephard1991characteristic}, we have
\begin{equation*}
\tilde{\mathbb{P}}(S_T\leq E)=\frac{1}{2}-\frac{1}{\pi}\int_0^{+\infty}\rm{Real}\bigg(\frac{e^{-\hat{i}\eta\log(K)}f_2(\eta,t)}{j\eta}\bigg)d\eta,
\end{equation*}
with $f_2(\eta,t)$ denoting the characteristic function of $Y_T$ under the measure $Q_1$ and can be obtained as 
\begin{eqnarray*}
f_2(\eta,T)&=&\mathbb{E}_t^{Q_1}\big(e^{\hat{i}\eta Y_T}\big)\\
		\nonumber &=& \mathbb{E}_t^{Q}\big(e^{-r(T-t)+Y_T-Y_t}e^{\hat{i}\eta Y_T}\big)=e^{-r(T-t)-Y_t}\mathbb{E}_t^{Q}\big(e^{(1+\hat{i}\eta) Y_T}\big)=e^{-r(T-t)-Y_t}f_1(\eta-\hat{i},t).
	\end{eqnarray*}
In order to find the price of put option, we need to determine the characteristic function $f_1(\eta,t)$. We now find an analytic expression for $f_1(\eta,t)$ by applying the Feynman-Kac theorem. Applying the tower property of expectation, we have:
\begin{equation*}
f_1(\eta, t) = \mathbb{E}_t\left[e^{\hat{i}\eta Y_T}\right]= \mathbb{E}_t\left[\mathbb{E}_t\left[e^{\hat{i}\eta Y_T} \mid X_T\right] \mid X_t\right].
\end{equation*}
where $X_t$ is the Markov chain governing the regime changes. We adopt a two step procedure to determine $f_1(\eta, t)$. Specifically, we first solve the inner conditional expectation $\mathbb{E}_t\left[e^{\hat{i}\eta Y_T} \mid X_T\right]$ which is a deterministic function of the Markov chain, and then we solve for the outer conditional expectation. Denote the inner expectation by $h(\eta,t\mid X_T)$. Using Feynman-Kac theorem, we have the following PDE governing $h(\eta,t\mid X_T)$,
\begin{equation*}
\frac{\partial h}{\partial t} + \frac{1}{2} \sigma_t^2 \frac{\partial^2 h}{\partial y^2}+ \left(r - \frac{1}{2} \sigma_t^2\right) \frac{\partial h}{\partial y}= 0,
\end{equation*}
with boundary condition:
\begin{equation*}
h(\eta, t\mid X_T)|_{t=T} =e^{\hat{i}\eta y_T}.
\end{equation*}
We assume that \( h(\eta,t \mid X_T)\) takes the following affine form
\begin{equation*}
h(\eta,t \mid X_T) = e^{C(\eta, \tau) + \hat{i}\eta y},
\end{equation*}
with \(\tau = T - t \). Substituting this into the PDE governing $h$ yields the following ordinary differential equations (ODEs)
\begin{equation*}
\frac{\partial C}{\partial \tau} = \frac{1}{2} \sigma_t^2 \eta^2 + (r-\frac{1}{2}\sigma_t^2) \hat{i} \eta,
\end{equation*}
with $C(\eta,0)=0$. The above ODE can be solved analytically and the solution is given by
\begin{equation*}
C(\eta,\tau) =\int_t^T \langle (r-\frac{1}{2}\sigma_s^2) \hat{i} \eta-\frac{1}{2} \sigma_s^2 \eta^2, X_s \rangle \, ds.
\end{equation*}
Therefore, the inner conditional expectation $h(\eta,t\mid X_T)$ is given by
\begin{equation*}
h(\eta,t\mid X_T)=e^{\int_t^T \langle (r-\frac{1}{2}\sigma_s^2) \hat{i} \eta-\frac{1}{2} \sigma_s^2 \eta^2, X_s \rangle \, ds+\hat{i}\eta y}.    
\end{equation*}
Next, we need to find expectation of $h$ with respect to the Markov chain, i.e.,
\begin{equation*}
f_1(\eta, t) = \mathbb{E}_t\left[h(\eta,t\mid X_T) \mid X_t\right]=\mathbb{E}_t\bigg(e^{\int_t^T \langle (r-\frac{1}{2}\sigma_s^2) \hat{i} \eta-\frac{1}{2} \sigma_s^2 \eta^2, X_s \rangle \, ds+\hat{i}\eta y}\mid X_t\bigg),
\end{equation*}
which gives
\begin{equation*}
f_1(\eta, t) =e^{\hat{i}\eta y+\hat{i}r\eta (T-t)} \mathbb{E}_t\bigg(e^{\int_t^T \langle (-\frac{1}{2}\sigma_s^2) \hat{i} \eta-\frac{1}{2} \sigma_s^2 \eta^2, X_s \rangle \, ds}\mid X_t\bigg).
\end{equation*}
Following \cite{elliott2013pricing},  
\begin{equation*}
\mathbb{E}_t\bigg(e^{\int_t^T \langle (r-\frac{1}{2}\sigma_s^2) \hat{i} \eta-\frac{1}{2} \sigma_s^2 \eta^2, X_s \rangle \, ds}\mid X_t\bigg)= \langle e^{M X_t}, I \rangle,
\end{equation*}
where \( X_t \in \{ e_1,e_2 \} \), and \( I = (1, 1)^T \) and the matrix \( M \) is given by:
\begin{equation*}
M = \int_t^T \left[ Q^T + \text{diag}\left(\frac{1}{2} \sigma_s^2 (-\hat{i}\eta-\eta^2)\right) \right] \, ds,
\end{equation*}
with \( Q \) being the transition rate matrix:
\begin{equation*}
Q = \begin{bmatrix}
    -\lambda_{12} & \lambda_{12} \\
    \lambda_{21} & -\lambda_{21}
\end{bmatrix}.
\end{equation*}
Solving the integration in the expression for \( M \) leads to the analytical formula for \( M \) as:
\begin{equation*}
M = \begin{bmatrix}
    -\lambda_{12}\tau + \frac{1}{2} \sigma_1^2 (-j\eta-\eta^2)\tau & \lambda_{21}\tau \\
    \lambda_{12}\tau & -\lambda_{21}\tau + \frac{1}{2} \sigma_2^2 (-j\eta-\eta^2)\tau
\end{bmatrix}.
\end{equation*}

\subsection*{\textbf{Regime switching Heston stochastic volatility model}}
Consider the Heston stochastic volatility model 
\begin{equation*}
    \begin{split}
        dS_t &= rS_tdt+\sqrt{v_t}S_tdW^1_t\\
        dv_t &=\kappa(\gamma-v_t)dt+\sigma_{X_t}\sqrt{v_t}dW^2_t
    \end{split}
\end{equation*}
The Monte Carlo method for pricing European put options using the above SDE is as follows
\begin{itemize}
    \item Discretise the above equations using the Euler Maruyama scheme
    \begin{equation*}
        \begin{split}
            S_{i+1} &= rS_i\delta t+\sqrt{v_i}S_i\Delta W_i^1\\
            v_{i+1} &= \kappa(\gamma-v_i)\delta t+\sigma_{X_i}\max\{\sqrt{v_i}, 0\}\Delta W_i^2
        \end{split}
    \end{equation*}
    where $X_i \in \{1, 2\}$.
    \item Generate N different realisations of $\{S_t\}_{0\leq t\leq T_{\rm{max}}}$ and $\{v_t\}_{0\leq t\leq T_{\rm{max}}}$ upto time $T_{\rm{max}}$ using the initial price $S_0$, initial volatility $v_0$ and initial regime $l\in\{1, 2\}$. Let $S^{jl} = \{S^{jl}_t|0\leq t\leq T_{\rm{max}}\}_{j=1}^N$, be the $N$ different realisations of stock prices with initial regime $l$.
    \item Using the generated paths evaluate $N$ different realisations of the payoff, i.e. $P_{t_0}^{jl}=\max\{0, E-S^{jl}_{t_0}\}$ at time $T=t_0$.
    \item The put option price in $l^{\rm{th}}$ regime with time to maturity $\tau=T-t=t_0$ is calculated as the average discounted payoff across all the $N$ realisations of the payoffs, i.e.
    \begin{equation*}
        P^l(t_0) = e^{-rt_0}\frac{1}{N}\sum_{j=1}^NP^{lj}_{t_0}.
    \end{equation*}
\end{itemize}

\end{document}